\documentclass[twocolumn,showpacs,preprintnumbers,amsmath,amssymb,superscriptaddress]{revtex4}

\newcommand{\beq}{\begin{equation}}
\newcommand{\eeq}{\end{equation}}
\def\beqa{\begin{eqnarray}}
\def\eeqa{\end{eqnarray}}
\def\hmpc{h^{-1}\,{\rm Mpc}}
\def\hkpc{h^{-1}\,{\rm kpc}}
\def\lcdm{\Lambda{\rm CDM}}

\usepackage{graphicx}
\usepackage{dcolumn}
\usepackage{bm}
\usepackage{color}

\def\apjl{ApJL}                
\def\apjs{ApJS}               
\def\aap{A\&A}                

\def\hmpc{~h^{-1} {\rm Mpc}}
\def\hkpc{~h^{-1} {\rm kpc}}

\begin{document}

\title{Hierarchy of N-point functions in the $\lcdm$ and ReBEL cosmologies}

\author{Wojciech A. Hellwing}
\email{pchela@camk.edu.pl}
\affiliation{Nicolaus Copernicus Astronomical Center, Bartycka 18, 00-719 Warsaw, Poland}
\affiliation{Interdisciplinary Center of Mathematical and Computational Modeling, Warsaw University, Poland}
\author{Roman Juszkiewicz}
\email{roman@camk.edu.pl}
\affiliation{Nicolaus Copernicus Astronomical Center, Bartycka 18, 00-719 Warsaw, Poland}
\affiliation{Institute of Astronomy, University of Zielona G\'ora, ul. Lubuska 2, Zielona G\'ora, Poland}
\author{Rien van de Weygaert}
\email{weygaert@astro.rug.nl}
\affiliation{Kapteyn Astronomical Institute, University of Groningen, P.O. Box 800, 9725LB Groningen, the Netherlands}

\date{\today}

\begin{abstract}
In this work we investigate higher order statistics for the $\lcdm$ and ReBEL scalar-interacting dark matter models by 
analyzing $180\hmpc$ dark matter N-body simulation ensembles. The N-point correlation functions and the related hierarchical 
amplitudes, such as skewness and kurtosis, are computed using the Count-In-Cells method. Our studies demonstrate that 
the hierarchical amplitudes $S_n$ of the scalar-interacting dark matter model significantly deviate from the values in the 
$\lcdm$ cosmology on scales comparable and smaller then the screening length $r_s$ of a given scalar-interacting model. 
The corresponding additional forces that enhance the total attractive force exerted on dark matter particles at 
galaxy scales lowers the values of the hierarchical amplitudes $S_n$. We conclude that hypothetical additional exotic 
interactions in the dark matter sector should leave detectable markers in the higher-order correlation statistics of 
the density field. We focussed in detail on the redshift evolution of the dark matter field's skewness and kurtosis. 
From this investigation we find that the deviations from the canonical $\lcdm$ model introduced by the presence of 
the ``fifth'' force attain a maximum value at redshifts $0.5<z<2$. We therefore conclude that moderate redshift data 
are better suited for setting observational constraints on the investigated ReBEL models. 
\end{abstract}


\maketitle

\section{\label{introduction}Introduction}
The standard hierarchical structure formation scenario assumes that the distribution of mass in the universe has grown  
out of primordial post-inflationary Gaussian density and velocity perturbations via gravitational instability. The resulting 
large-scale structures can be described in a statistical way. The two-point and higher order correlation functions are the most 
widely studied measures. For the standard cold dark matter paradigm -- which now is a part of the commonly accepted $\lcdm$ 
model -- these have been studied analytically \citep[e.g.][]{1980Peebles,Juszkiewicz1993,ber1992}, as well as numerically on 
the basis of N-body cosmological simulations \citep[e.g.][]{npoint_omega_cdm,skew_nongaussian,cic_nbody1,cic_nbody2}. 

Here we concentrate our study on a modified dark matter model that includes long-range scalar interactions between DM particles. 
We focus on the phenomenological model of such a long range ``fifth'' DM force proposed in a study by Farrar, Gubser and Peebles 
\cite{FarrarPeebles,PeeblesIDMDE,GP1,GP2,Farrar2007,Brookfield}. We follow Keselman \textit{et al.} in dubbing this long-range scalar 
interaction model as ReBEL, daRK Breaking of Equivalence principLe. This model was proposed as a possible remedy for some of the 
$\lcdm$ problems, which relate mostly to galaxy scales. For an excellent discussion of the motivation behind the long-range 
scalar-interacting model we refer to papers by Peebles \cite{PeeblesVoid,peeb2} and a recent review by Peebles \& Nusser \cite{peeb1}.
Over the past few years, the ReBEL model has been extended and explored in a range of studies 
\cite{NGP,LRSI1,LRSI2,Rebel,Rebel2,reb_sat1,reb_sat2}. These studies have revealed its potential on the basis of promising results. A 
variety of similar models have also been studied, mostly by means of N-body simulations \cite{Coupled_DE1,Coupled_DE2,Coupled_DE3,Coupled_DE4,Coupled_DE5,Coupled_DE6,Coupled_DE7,COupled_DE8}.
There are also additional observational arguments in favour of the ``fifth force'' in the 
dark matter sector, recently forwarded by \cite{lee}.

In this paper we study the hierarchy of N-point correlation functions of the scalar-interacting DM ReBEL model. In principle, these 
can be used to infer observational constraints on the free parameters of the model. This is not an entirely trivial affair, since the 
comparison of the results with observations is somewhat complicated by a few factors: (1) galaxies do not necessarily trace the mass 
(biasing) and (2) in the ReBEL model the baryonic matter is insensitive to the extra scalar forces. Nonetheless, we expect that 
the information content of the higher order correlation functions is sufficient to distinguish between the standard DM and 
scalar-interacting DM paradigms.

To study the high order correlations patterns of the DM density field we use cosmological N-body simulations. The scale and resolution 
of the simulations are designed such that they are perfectly suited for our purpose, i.e. they address the highly nonlinear evolution 
at scales smaller than $\approx 10 \hmpc$. For the purpose of distinguishing between cosmologies these scales are particularly 
useful, since (1) the expected deviation of the ReBEL model from the canonical $\lcdm$ is maximal at these small fully 
nonlinear scales \cite{NGP,LRSI1}, and (2) nearly all detailed observations, except for the largest galaxy catalogs, relate to the small 
or intermediate scales. 

This paper is organized as follows: in section \ref{sec-model} we describe scalar-interacting DM ReBEL model, followed in 
section~\ref{sec-numerics} by the description of the numerical modelling. Section~\ref{sec-method} covers the issues 
related to the Counts-In-Cells method to sample the N-point correlation functions. The results of our study are presented 
in section~\ref{sec-results}, followed by the conclusions in section~\ref{sec-conclusions}.

\section{Scalar-interacting Dark Matter Model}
\label{sec-model}
Following our previous work \cite{LRSI1} we study the model of the ReBEL long-range scalar interactions in the dark matter sector. 
In this scenario dark matter particles interact by means of an additional ``fifth'' force mediated by a massless scalar. The extra 
force term is long-range, even though it is dynamically screened by a sea of light particles coupled to the scalar field 
\cite{GP1,GP2}. The resulting effective gravitational potential between two DM particles has the form \cite{NGP}\,:
\beq
\label{eqn:lrsi_pot}
\Phi({\bf r}) = -\,{Gm\over r}\,\,g(x)\,\,,
\eeq
in which $G$ is Newton's constant and  
\beq
\label{eqn:g(x)}
g(x) \,=\, 1 + \beta\, e^{-x/r_s} \, .
\eeq
In this expression ${\bf r}$ and ${\bf x}$ are the particle separation in real and comoving space. The cosmological scale 
factor $a(t)$ at cosmological time $t$ is normalized to unity at the present epoch, $a(t_0)\;=\;1\;$. The model is specified 
by means of two parameters:
\begin{enumerate}
\item[$\bullet$] $\beta$:  strength parameter\\
The strength parameter $\beta$ is a dimensionless measure of the strength of the scalar interaction with respect to a pure Newtonian 
gravitational gravitational force: for $\beta = 1$ the ReBEL forces between two dark matter particles are of the same magnitude 
and strength as the Newtonian gravitational force. 
\item[$\bullet$] $r_s$: scale parameter\\
the comoving screening length in $\hmpc$, which remains constant in the comoving frame. 
\end{enumerate}
The total effective force between two dark matter particles of mass $m_1$ and $m_2$ is   
\beq
F_{DM} = -G{m_1\cdot m_2\over r^2}\left[1+\beta\left(1+{r\over r_s}\right)e^{-{r \over r_s}}\right]\,.
\label{eqn:lrsi_force}
\eeq
From this expression we may immediately infer that the regular Newtonian force is recovered at 
distances $r\gg r_s$, while for separations $r \leq r_s$ the force experienced by the dark matter 
particle will be enhanced or reduced with respect to the Newtonian force (depending on the sign 
of the strength parameter $\beta$). 

\section{Numerical simulations}
\label{sec-numerics}
\begin{table*}
\caption{\label{tab:sim_params} Parameters used in our ensembles of simulations. No. of realizations stands for the number of different realizations of the same initial $P(k)$, $\beta$ and $r_s$ are scalar-interactions parameters. $L_{box}$ denotes the size of the 
simulation box, $N_{part}$ the number of particles and $z_{ini}$ the initial redshift. The cosmological parameters are: 
$\Omega_m$ and $\Omega_{\Lambda}$, denoting the dimensionless density parameters of the matter and cosmological constant at 
redshift $z=0$, and $\sigma_8$, the amplitude of mass fluctuations in a $8\hmpc$ sphere, h is the present dimensionless Hubble parameter, $m_p$ is the particle mass, $\varepsilon$ marks the force resolution and $l$ denotes the mean interparticle separation.}
\medskip
\begin{tabular}{lcccccccccccccccc}
\hline
&&&&&&&&&&&&&&&&\\
ensemble & No. of rea-& & $\beta$ & $r_s$ & & $\Omega_m$ & $\Omega_{\Lambda}$ & h & $\sigma_8$ & & $L_{box}$ & $N_{part}$ & $z_{ini}$ & $m_p$ & $\varepsilon$ & $l$\\
 & lizations & & $[\hkpc]$ & $[\hmpc]$ & & & & $[100\,\frac{\textrm{km}}{\textrm{s}\cdot\textrm{Mpc}}]$ & & & $[\hmpc]$ & & & $[10^{10}M_{\odot}]$ & $[\hkpc]$  \\
 &&&&&&&&&&&&&&&&\\
\hline 
&&&&&&&&&&&&&&&&\\
1024SCDM & 10 && - & - && 1.0 & 0.0 & 0.5 & 1.0 && 1024 & $256^3$ & 35 & 1776.32 & 924 & 4 \\
&&&&&&&&&&&&&&&&\\
180LCDM & 8 && - & - && 0.3 & 0.7 & 0.7 & 0.8 && 180 & $256^3$ & 40 &  2.89 & 168 & 0.703 \\
360LCDM & 5 && - & - && 0.3 & 0.7 & 0.7 & 0.8 && 360 & $256^3$ & 30 &  23.155 & 168 & 1.4\\
512LCDM & 5 && - & - && 0.3 & 0.7 & 0.7 & 0.8 && 512 & $256^3$ & 30 & 66.612 & 280 & 2\\
&&&&&&&&&&&&&&&&\\
180B-05RS1 & 8 && -0.5 & 1 && 0.3 & 0.7 & 0.7 & 0.8 && 180 & $256^3$ & 40 & 2.89 & 168 & 0.703 \\
180B02RS1 & 8 && 0.2 & 1 && 0.3 & 0.7 & 0.7 & 0.8 && 180 & $256^3$ & 40 & 2.89 & 168 & 0.703 \\
512B02RS1 & 5 && 0.2 & 1 && 0.3 & 0.7 & 0.7 & 0.8 && 512 & $256^3$ & 30 & 66.612 & 280 & 2\\
180B1RS1 & 8 && 1 & 1 && 0.3 & 0.7 & 0.7 & 0.8 && 180 & $256^3$ & 40 & 2.89 & 168 & 0.703 \\
360B1RS1 & 5 && 1 & 1 && 0.3 & 0.7 & 0.7 & 0.8 && 360 & $256^3$ & 30 & 23.155 & 168 & 1.4\\
512B1RS1 & 5 && 1 & 1 && 0.3 & 0.7 & 0.7 & 0.8 && 512 & $256^3$ & 30 & 66.612 & 280 & 2\\
&&&&&&&&&&&&&&&&\\
180LCDMZ80 & 8 && - & - && 0.3 & 0.7 & 0.7 & 0.8 && 180 & $256^3$ & 80 & 2.89 & 16.8 & 0.703\\
180B1RS1Z80 & 8 && 1 & 1 && 0.3 & 0.7 & 0.7 & 0.8 && 180 & $256^3$ & 80 & 2.89 & 16.8 & 0.703\\
&&&&&&&&&&&&&&&&\\
256LCDMHR & 10\footnote{These simulations have only 1 realisation, we used 10 bootstrap resamplings to obtain the estimates of the mean and standard deviation.} && - & - && 0.3 & 0.7 & 0.7 & 0.8 && 256 & $512^3$ & 80 & 1.04 & 16.8 & 0.5\\
256B1RS1HR & $10^a$ && 1 & 1 && 0.3 & 0.7 & 0.7 & 0.8 && 256 & $512^3$ & 80 & 1.04 & 16.8 & 0.5\\
&&&&&&&&&&&&&&&&\\
\hline
\end{tabular}
\end{table*}

A series of N-body numerical experiments is used to trace and investigate the growth of the large-scale 
structure in various cosmological scenarios. Part of the simulations concern the canonical ``concordance'' 
$\Lambda$CDM cosmology. Most simulations involve different versions of ReBEL cosmologies. In addition, 
10 large-scale SCDM cosmology simulations are invoked for testing purposes. 

A listing of the parameters and settings of the ensembles of the simulations is provided by 
Table~\ref{tab:sim_params}. Simulations of the concordance $\Lambda$CDM cosmology are labeled with LCDM, 
while the ReBEL ones are labeled with $B$ and $RS$ and related parameters indicating the $\beta$ and $r_s$ 
parameters of the scalar-interacting dark matter. The digits at the beginning of each label relate to the 
size of the simulation box. In addition to the specific scenario characteristics - such as 
$\Omega_m$, $\Omega_\Lambda$, Hubble parameter $H$, $\sigma_8$ and ReBel Parameters $\beta$ and 
$r_s$ - the simulations differ in terms of the simulation box size $L_{box}$, number of particles 
$N_{part}$, force resolution and initial redshift $z_{ini}$. 

The simulations in a $180\hmpc$ box form the core of our study, with the simulations in larger 
boxes kept for additional analysis. With the exception of the 256LCDMHR and 256B1RS1HR 
ensembles, all numerical simulations contain $256^3$ dark matter particles to sample the theoretical 
continuum density dark matter field. Simulations 256LCDMHR and 256B1RS1HR, consisting of $512^3$ dark matter 
particles, and simulation ensembles 180LCDMZ80 and 180B1RS1Z80 have a higher force resolution,  
$\varepsilon=16.8\hkpc$. These simulations are used to study the transients and resolution effects. 

For each configuration of simulation parameters we generate an ensemble of 5-10 different simulations. This 
enables us to get an estimate of the cosmic variance introduced by the finite simulation box sizes. Each of 
the ensemble realizations is based on the same amplitude of the density field's Fourier components, dictated 
by the power spectrum, while differing in terms of the corresponding random phases. 

The initial density and velocity fluctuation field in all simulations are characterized by a 
cold dark matter spectrum. To generate the initial conditions we use the 
\verb#PMcode# by Klypin \& Holtzman \cite{PMcode}, in conjunction with transfer functions computed 
using the \verb#cmbfast# code by Seljak \& Zaldarriaga \cite{cmbfast}. With the exception of the 
Standard Cold Dark Matter SCDM model, all $\Lambda$CDM and ReBEL models start from an initial 
density field with a canonical $\lcdm$ power spectrum normalized to a linearly extrapolated 
density variance $\sigma_8 = 0.8$ at redshift $z=0$ within a sphere of comoving tophat radius 
$R_{TH}=8\hmpc$. 

The 1024SCDM ensemble traces growth of structure in the Standard Cold Dark Matter (SCDM) model. Each 
of the 10 realizations are contained within a $1024\hmpc$ cubical box. Even though currently the 
SCDM model is very strongly disfavored by all astronomical data \citep[e.g.][]{SCDM1,SCDM2,SCDM3,SCDM_GA,
FeldmanOmega,acceleration1,acceleration2}, we use it as reference point and for testing purposes on the 
grounds that over the past decades it has been studied in great detail \citep[e.g.][]{npoint_omega_cdm,
cic_ana,1980Peebles,White_SCDM,SCDM4}.

To evolve the particle distribution from the initial scale factor to the present time we use the \verb#Gadget2# 
Tree-Particle-Mesh code by Volker Springel \cite{Gadget2}, which we specifically modified to be able to follow the 
particle distribution in ReBEL force fields. The modifications allow the code to handle the long-range 
scalar-interacting dark matter interactions (eqn.~\ref{eqn:lrsi_pot}-\ref{eqn:lrsi_force}). The detailed 
description of this modification may be found in our earlier work \cite{LRSI1}. Of all simulations, we saved 
particle positions and velocities at redshifts $z=5, 2, 1, 0.5$ and $0$. The end product is a catalog of 
redshift-dependent \textit{snapshots}. 

In a simulation with a (comoving) box size of $180\hmpc$, a dark matter particle has a mass of 
$2.89\times 10^{10}h^{-1}M_{\odot}$. In this case, a typical galaxy halo will contain roughly 
a hundred dark matter particles. This number is too small to reliably sample any relevant physical 
quantities of a galaxy halo. However, it is sufficient to reliably trace the non-linear evolution 
of the dark matter density field down to scales relevant for galaxy formation. 

\section{Moments of counts-in-cells}
\label{sec-method}
Assuming the applicability of the fair-sample hypothesis \footnote{the fair-sample hypothesis states that the ensemble 
average of a stochastic perturbation field is equal to the average over a large number of sampling volumes in the 
Universe}, the volume-averaged $J$-point correlation function can be expressed as 
\beq
\label{volume-aver-nfunction}
\bar{\xi_J}\,=\,V_W^{-J}\int_S d\mathbf{x_1}...d\mathbf{x_J}W(\mathbf{x_1})...W(\mathbf{x_J})\xi_J(\mathbf{x_1},...,\mathbf{x_J})\,,
\eeq
where $\mathbf{x_i}$ is the comoving separation vector, $W(\mathbf{x})$ is a window function with volume 
\beq
V_W\,=\,\int_S d\mathbf{x}\,W(\mathbf{x})\,,
\eeq
and the integral covers the entire volume $S$. Because of the fair-sample hypothesis, $\bar{\xi_J}$ does not depend 
on the location $\mathbf{x}$ and is a function of the window volume $V_W$ only \cite{1980Peebles}\,. 

\subsection{Connected Moments}
There is a range of options concerning fast and accurate methods for measuring the N-point correlation functions 
of a DM density field sampled by a discrete set of $N$ particles. Our analysis is based on the moments of the 
distribution of counts-in-cells (hereafter CIC) \cite{1980Peebles,GaztanagaAPM94,cic_ana,BCGS_book}. The counts 
define a discrete sample of the density distribution. Sampling the density field by $C$ spherical cells, the 
$J$-th central moment of the cell counts is defined by
\beq
\label{eqn-j-moment}
m_j(R) = {1\over C}\sum_{i=1}^{C}(N_i-\tilde{N})^J\,\,,
\eeq
where $R$ is the comoving cell radius, $N_i$ the number of particles found in a $i$-th cell and $\tilde{N}$ the mean 
number of particles in cells of radius $R$. Following Gazta\~naga\cite{GaztanagaAPM94}, the connected moments $\mu_j$ 
of the counts may then be written as,  
\beqa
\mu_2 &=& m_2,\cr
\mu_3 &=& m_3,\cr
\mu_4 &=& m_4 - 3m_2^2,\cr
\mu_5 &=& m_5 - 10m_3m_2,\cr
\mu_6 &=& m_6 - 15m_4m_2 - 10m_3^2+30m_2^3,\cr
\mu_7 &=& m_7 -21m_5m_2 - 35m_4m_3 + 210m_3m_2^2,\cr
\mu_8 &=& m_8 -28m_6m_2 - 56m_5m_3 - 35m_4^2\cr
&+& 420m_4m_2^2 + 560m_3^2m_2 - 630m_2^4,\cr
\mu_9 &=& m_9 - 36m_7m_2 - 84m_6m_3 - 126m_5m_4 + 756m_5m_2^2\cr
&+& 2520m_4m_3m_2 + 560m_3^3 - 7560m_3^2m_2^3.
\label{eqn-connected-moments}
\eeqa

\medskip
The volume-averaged correlation functions $\bar{\xi}_J$ can be computed by dividing the equations for 
the connected moments by $\tilde{N}^J$, 
\beq
\label{counts-nfunctions}
\bar{\xi}_J = \mu_j \tilde{N}^{-J} \,.
\eeq

\subsection{Shot-Noise effects}
Due to the discrete nature of a finite particle distribution, equation~\ref{counts-nfunctions} is a good estimator of 
$\bar{\xi_J}$ only for scales where the fluid limit holds. This is satisfied if $\tilde{N}\gg 1$. For small values of 
$\tilde{N}$ or, more adequately, for scales comparable with the mean inter-particle separation, the 
factor $\mu_j \tilde{N}^{-J}$  will be dominated by shot noise. 

To correct for the shot noise effects, we use the method developed by Gazta\~naga \citep[see][]{GaztanagaAPM94}. The 
method use the moment generating function of the Poisson model to calculate the net contribution by discrete noise. 
By including this information, one may infer expressions for the shot-noise corrected connected moments $k_J$:
\beqa
k_2 &=& \mu_2 - \tilde{N},\cr
k_3 &=& \mu_3 - 3k_2 - \tilde{N}, \cr
k_4 &=& \mu_4 - 7k_2 - 6k_3 -\tilde{N}, \cr
k_5 &=& \mu_5 - 15k_2 - 25k_3 - 10k_4 - \tilde{N}, \cr
k_6 &=& \mu_6 - 31k_2 - 90k_3 - 65k_4 - 15k_5 - \tilde{N}, \cr
k_7 &=& \mu_7 - 63k_2 - 301k_3 - 350k_4 - 140k_5 - 21k_6 - \tilde{N}, \cr
k_8 &=& \mu_8 - 127k_2 - 966k_3 - 1701k_4 - 1050k_5 - 266k_6 \cr
 &-& 28k_7 - \tilde{N},\cr
k_9 &=& \mu_9 - 255k_2 - 3025k_3 - 7770k_4 - 6951k_5 - 2646k_6\cr
&-& 462k_7 - 36k_8 - \tilde{N}.
\label{corrected-connected-moments}
\eeqa
Finally the corrected volume-averaged $J$-th point correlation functions of DM density field can be written as
\beq
\label{cor-npoint-f}
\bar{\xi}_J\,=\,k_J\tilde{N}^{-J}.
\eeq

We use relations described above to compute  $\bar{\xi}_J$'s up to $J=9$ from the particle distributions of our N-body cosmological simulations.

\subsection{Sampling and errors}
Because the computational cost of counting the content of cells increases with volume, we adjust the number 
of spherical cells used for the counts-in-cells analysis to the comoving cell radius $R$. We require the total number 
of sampling spheres to be in the range $10^5\leq C\leq 10^6$. For the smallest scales we take $C=10^6$, while for the 
largest scales the minimum number of cells is $10^5$. Within this range, the number of cells used to sample the moments, $C(R)$, 
scales according to  
\beq
\label{eqn:sampling}
C(R)\propto \left({L\over R}\right)^3\;\;,
\eeq
where $L$ is the comoving simulation box width. This scaling implies the number of counted points as function of 
scale $R$ to remain comparable. 

Constraining the number of sampling cells is a trade-off between the requirement of keeping the sampling errors as low 
as possible and limits on the computational time. Because the sampling error connected with the finite number of 
cells $C$ scales like $C^{-1}$ \cite{cosmic_error}, the decreasing number of cells at larger radii $R$ leads to a 
corresponding growth of the intrinsic error. 

In this paper we adopt the standard deviation on the mean of the $J$-point correlation function, determined 
from its estimated values $\xi^i_J$ in the various realizations $i$ ($i=1,\ldots,M$) within a simulation ensemble 
(see ~\ref{tab:sim_params}),  
\beq
\langle\bar{\xi}_J\rangle\,=\,{\displaystyle 1 \over \displaystyle M-1}\,\sum_{i=1}^M\,\xi^i_J\,
\eeq
as a measure for the variability and error $\sigma_{\xi_J}$ in the estimate for the correlation function $\xi_J$, 
\beqa
\label{eqn:stdev}
\sigma_{\xi_J}\,=\,\sqrt{\textrm{Var}[\bar{\xi}_J]} = \sqrt{{\displaystyle 1 \over \displaystyle M-1}\,{\sum_{i=1}^{M}(\bar{\xi}^i_J-\langle\bar{\xi}_J\rangle)^2}}\;\;,
\eeqa
The standard deviation of an ensemble obtained by averaging over its realizations concerns a conservative estimate of errors. The 
sampling variance is larger for different realizations within an ensemble than for measurement errors associated with the 
finite number of the sampling cells \cite{cic_ana}.

\section{Testing the Counts-in-Cells method}
We test our implementation of the CIC method by probing its performance with respect to its 
estimates of the two-point correlation function $\xi_2$ and the three-point correlation function $\xi_3$.

\subsection{Variance and 2nd order moment}
The second order moment is widely used to characterize the rms fluctuation of the matter density field on a given scale,
\beq
\label{sigma-xi}
\sigma^2(R)\,=\,\bar{\xi_2}(R)\,.
\eeq
where the scale $R$ is the comoving radius of the applied window function $W$. 

There are two routes towards determining this factor. The first estimate of $\bar{\xi_2}(R)$ is yielded by 
the counts-in-cells formalism. Following the Gazta\~naga formalism, CIC leads to the estimate (eqn.~\ref{cor-npoint-f})
\beq
{\widehat {\bar{\xi}}_2}[CIC](R)\,\equiv\,k_2\tilde{N}(R)^{-2}.
\label{varcic}
\eeq
where $\tilde{N}(R)$ is the number of particles in spherical cells of radius $R$. 

A second estimate of $\sigma(R)$ is based on the power spectrum $P(k)$ of the dark matter density field in the simulations. 
In theory, the variance follows directly from the power spectrum of density fluctuations $P(k)$, via the integral 
over the comoving wave number $k$,
\beq
\label{sigma-pk}
\bar{\xi_2}(R)\,=\,\sigma^2(R)\,=\,\int_0^\infty {\displaystyle dk \over \displaystyle 2 \pi^2}\,\,\, k^2 P(k){\hat W}^2(kR)\,.
\eeq
With our analysis being based on counts-in-cells in spherical volumes of radius $R$, the natural window function is the 
spherical tophat function. 

In the remainder, the spherical tophat function is used as window function. In Fourier space, the top-hat 
window function is specified by 
\beq
\label{top-hat}
{\hat W}_{TH}(kR)\,=\,3\,{\sin(kR)-kR\cos(kR)\over(kR)^3}.
\eeq
As a result of the discrete nature of the particles set and the finite size of the simulation box, the particle simulation 
cannot probe the density perturbations on scales larger than the simulation box length $L$ and smaller than the mean 
particle separation,
\beq
l\,\propto\,N/L^3\,.
\eeq
(for a simulation of $N$ particles in a box of length $L$). For a proper comparison with the CIC inferred variance, 
the corresponding density field estimate integral in equation~\ref{sigma-pk} is evaluated in between proper integral 
boundaries. The lower limit is the fundamental mode $k_L$, while the Nyquist frequency $k_{Nyq}$ represents 
the upper limit. For a box of size $L$, these are 
\beq
\label{k-limits}
k_L ={2\pi\over L},\qquad k_{Nyq} = k_L{N^{1/3}\over 2}\,,
\eeq
where we presume that the number of grid cells on which we have sample the initial density field is equal to the 
number of particles $N$. Hence, the power spectrum variance estimate is given by 
\beq
{\widehat {\bar{\xi}}_2}[Pk](R)\,\equiv\,\int_{k_L}^{k_{Nyq}} 
{\displaystyle dk \over \displaystyle 2 \pi^2}\,\,\, k^2 P(k)\,{\hat W}_{TH}^2(kR)\,.
\label{sigma-pk-pk}
\eeq

For all simulation runs (see table~\ref{tab:sim_params}), we have computed the nonlinear power spectra directly from 
the resulting simulation particle distributions \footnote{the nonlinear power spectrum is directly derived from the 
dark matter density field obtained from the simulation, while the linear (extrapolated) power spectrum is the primordial 
power spectrum multiplied by the appropriate linear density growth factor}. The integral in equation~\ref{sigma-pk-pk} is 
calculated from the computed nonlinear power spectra for a limited set of ensembles, those of 1024SCDM, 180LCDM, 
180B-05RS1, 180B02RS1, 180B1RS1 (see table~\ref{tab:sim_params}). 

\begin{figure*}
\includegraphics[width=0.5\textwidth,angle=-90]{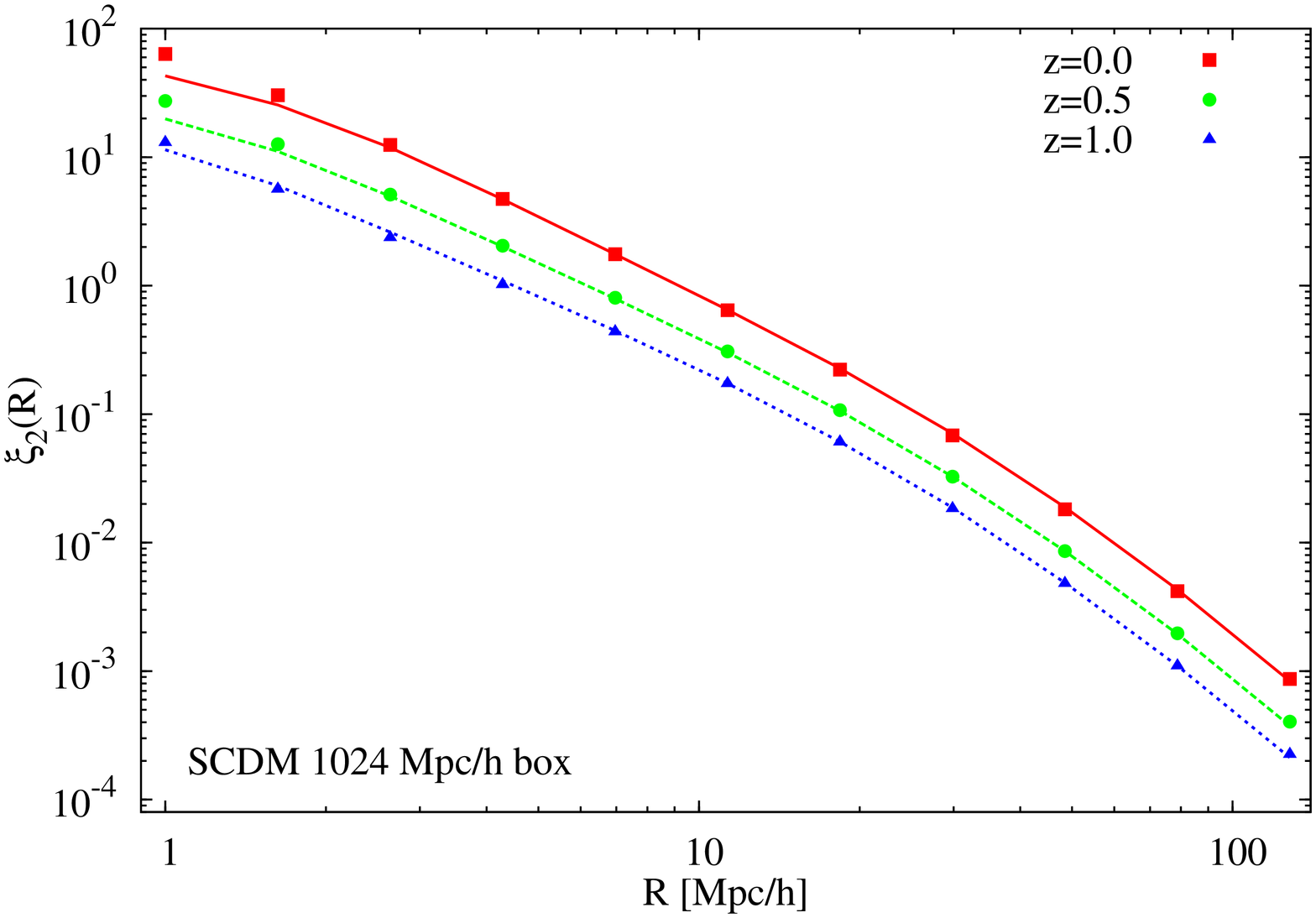}
\caption{Variance Estimators. Comparison of the CIC estimator ${\widehat {\bar{\xi}}_2}[CIC]$ and the power spectrum 
estimator ${\widehat {\bar{\xi}}_2}[Pk]$ of the variance as a function of scale $R_W$, based on the ensemble of 1024SCDM 
simulations. The variance is determined for three redshifts: $z=0$, $z=0.5$ and $z=1.0$. 
The symbols represent the CIC variance estimates. Filled squares: $z=0$, circles: $z=0.5$, triangles: $z=1.$. The continous 
lines indicate the power spectrum integral estimates. Solid line: $z=0$, dashed line: $z=0.5$, dotted line: $z=1$.}
\label{fig:xi2_test_scdm}
\bigskip
\includegraphics[width=0.5\textwidth,angle=-90]{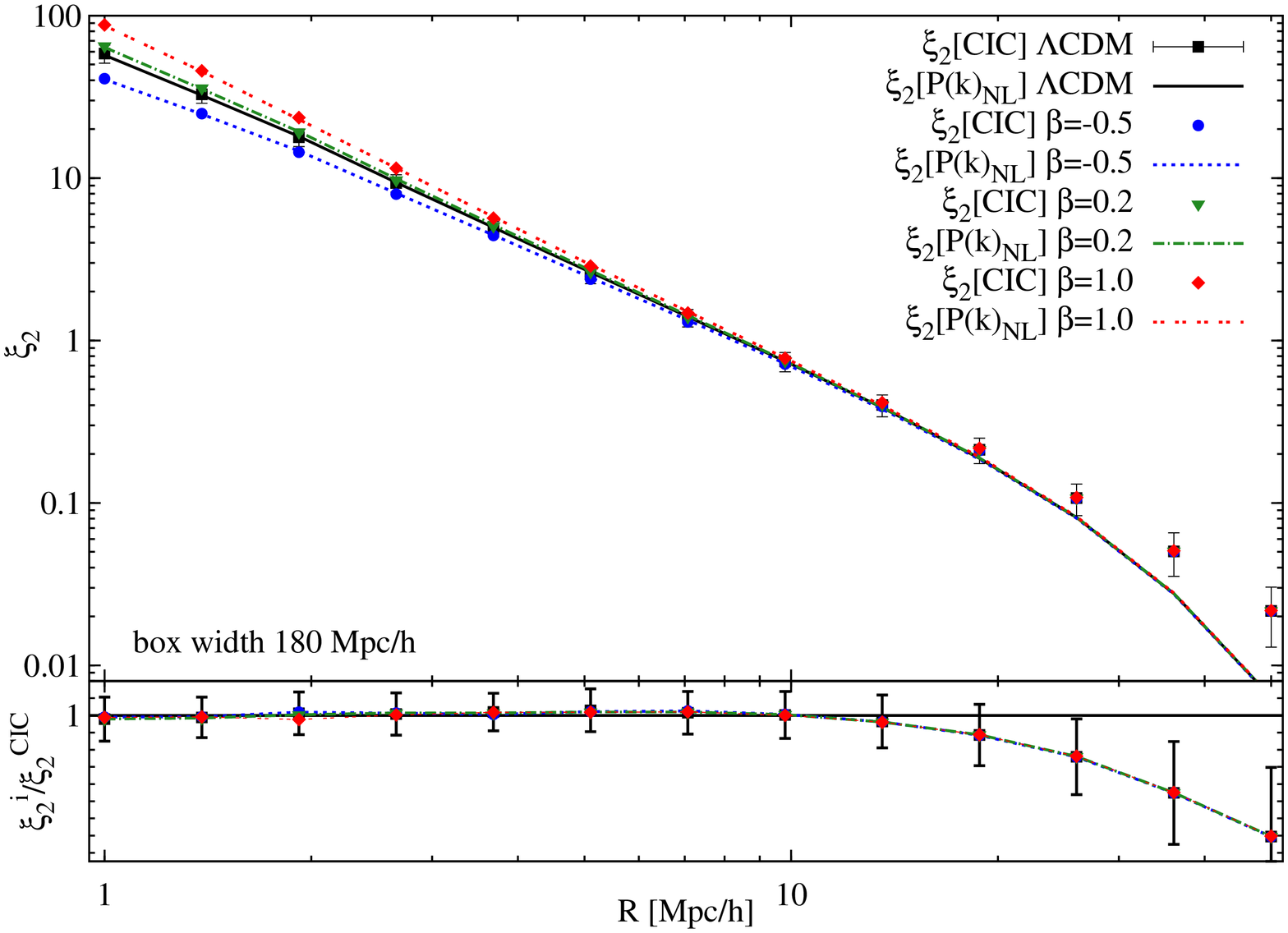}
\caption{Comparison between the CIC estimator ${\widehat {\bar{\xi}}_2}[CIC]$ and the power spectrum integral 
estimator ${\widehat {\bar{\xi}}_2}[Pk]$ of the variance of the density field, as a function of scale $R_W$. The 
panel shows a comparison between the two estimators for simulation ensembles of four different cosmologies. These 
concern the LCDM cosmology and three different ReBEL cosmologies. All simulations have a box size of $180\hmpc$. 
The power spectrum estimates are represented by continuous lines, the CIC estimates by corresponding symbols. The 
ensembles are: 1) 180LCDM - LCDM cosmology - solid line - square; 2) 180B-05RS1 - ReBEL cosmology with $\beta=-0.5$ - 
dotted line - circle; 3) 180B02RS1 - ReBEL cosmology with $\beta=0.2$ - dot-dashed line - triangle ; 4) 180B1RS1 - ReBEL 
cosmology - double-dotted line - diamond  with $\beta=1.0$. For clarity, we only show error bars for the 180LCDM 
ensemble. Top panel: regular plot of variance estimator vs. scale $R_W$. Bottom panel: Plot of estimator ratio 
${\widehat {\bar{\xi}}_2}[Pk]/{\widehat {\bar{\xi}}_2}[CIC]$.}
\label{fig:xi2_test_lcdm}
\end{figure*}

\subsubsection{Variance test} 
In Fig.~\ref{fig:xi2_test_scdm} we present a comparison between the two estimates of the variance $\sigma^2(R)$, 
i.e. between the estimate ${\widehat {\bar{\xi}}}_{2}[CIC]$ on the basis of the CIC method (eqn.~\ref{varcic}) and the 
estimate ${\widehat {\bar{\xi}}}_{2}[Pk]$ from the power spectrum integral (eqn.~\ref{sigma-pk}). For the ensemble of 
1024SCDM simulations, we determined the variance at three different redshifts, $z = 0, 0.5$ and $z=1.$. 
The diagram plots the resulting variance as a function of the scale $R$. The symbols ($z=0$: filled squares, $z=0.5$: 
circles, $z=1$: triangles) indicate the variance estimates on the basis of the CIC method. The continuous lines 
represent the variance determined from the power spectrum integral ($z=0$: solid, $z=0.5$: dashed, $z=1$: dotted). 

In the 1024SCDM simulation, the Nyquist frequency $k_{Nyq}\approx 0.785$ corresponds to $\sim 8\hmpc$. This means that  
the diagram in Fig.~\ref{fig:xi2_test_scdm} suffers from a substantial level of shotnoise contribution over the range between 
$1\hmpc<R<8\hmpc$. Nonetheless, the agreement between the two estimators is remarkably good down to a scale of 
$\approx 3\hmpc$, comparable to the mean inter particle separation in the 1024SCDM ensemble. 

\begin{figure*}
{\includegraphics[width=0.5\textwidth,angle=-90]{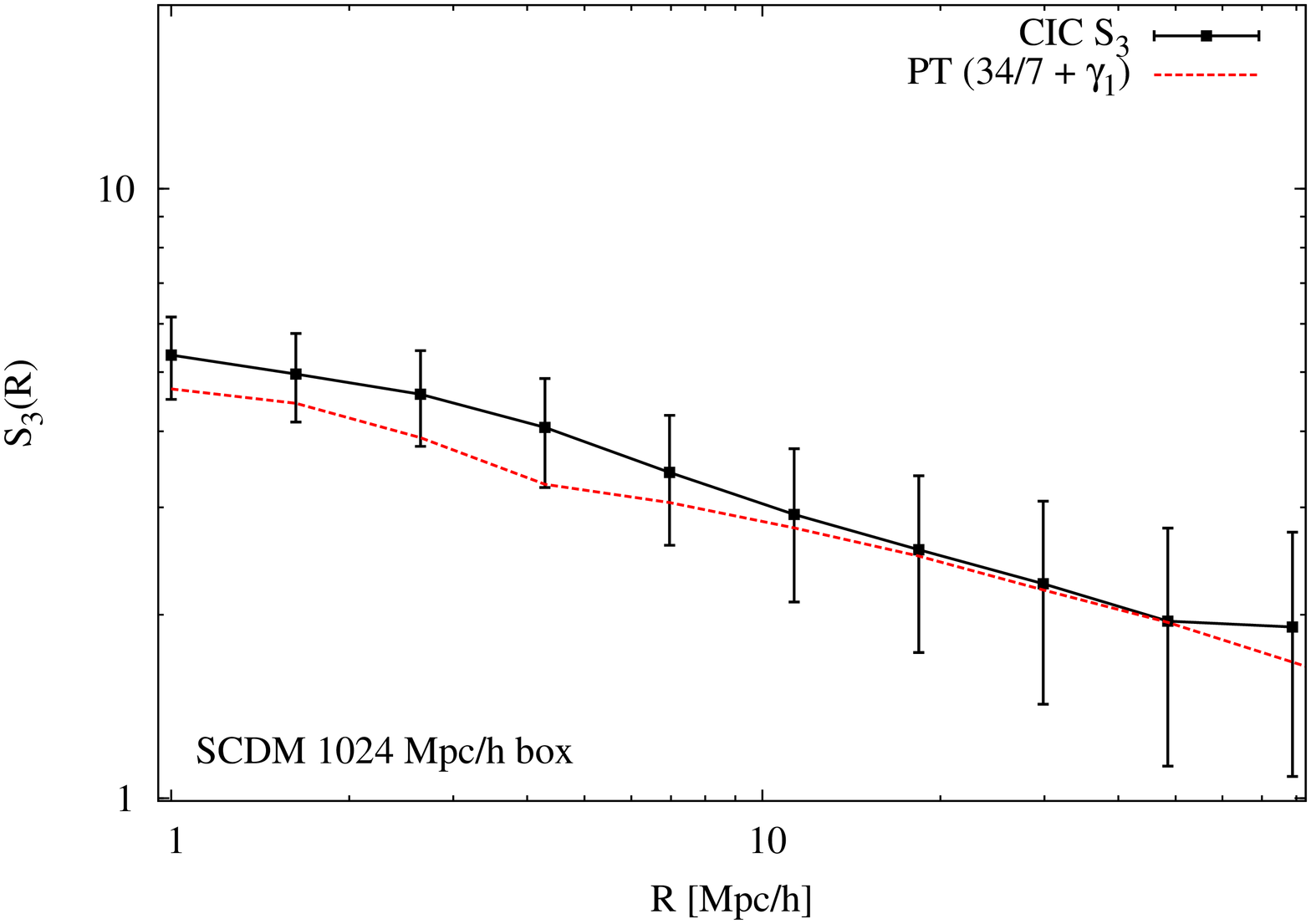}}
\medskip
\caption{The skewness $S_3$ measured for the SCDM simulation ensemble 1024SCDM. The solid line shows the CIC estimate of $S_3$ 
(eqn.~\ref{eqn:s3cic}). The dotted line shows the skewness estimated on the basis of perturbation theory (eqn.~\ref{eqn:s3_pt}). 
The error bars (only shown for the CIC) correspond to maximal standard deviation of the ensemble (see main text for the details). 
\label{fig:s3_test_scdm}}
\end{figure*}

\subsubsection{Variance Estimate \& Model Dependence}
To check whether the modified dynamics of the DM fluid in the ReBEL model affects the two variance estimators differently, we 
compare the resulting estimates for a range of different ReBEL models. 

The top panel of fig.~\ref{fig:xi2_test_lcdm} compares the two estimates at different scales $R_W$ for four different 
cosmologies: $\Lambda$CDM and a ReBEL model with strength parameter $\beta=-0.5$, a ReBEL model with $\beta=0.2$ and one 
with $\beta=1.0$. The lines represent the power spectrum estimate ${\widehat {\bar{\xi}}}_{2,Pk}$ (see legend). The CIC 
estimates of the variance are indicated by symbols of the same colour as the lines, listed in the legenda. The difference 
between the two estimates may be best appreciated from the bottom panel, which shows the ratio between the two 
estimators, ${\widehat {\bar{\xi}}_2}[Pk]/{\widehat {\bar{\xi}}_2}[CIC]$. 

Both panels clearly shows that for all four different cosmologies the two estimators agree very well for scales ranging 
from $\approx 10\hmpc$ down to the smallest scales that we analyzed, $\approx 1\hmpc$. On larger scales, from 
$\approx 20\hmpc$ (roughly $1/10$th of the box width), we see a marked disagreement between the two estimators. This 
difference rapidly increases towards larger scales, with the CIC estimate systematically increasing as a function of scale 
with respect to the power spectrum value. Nonetheless, the fact that the difference between the two estimates is identical 
for the different model ensembles, in terms of character and scale at which they start to diverge, indicates that the accuracy 
achieved by the CIC method is the same for each of the cosmologies. 

\subsection{Third order moment: the $S_3$ test}
The second order density field statistic, represented by the two-point correlation function, is not sufficient for 
characterizing the density field beyond the linear phase of structure evolution. Moving into the quasi-linear phase, we 
start to discern the gravitational contraction of overdense regions into sheetlike and filamentary patterns and 
compact dense haloes and the volume expansion of low density void regions. To be able to follow and characterize 
this process, we need to turn to the higher order moments of the density field.

To test the performance of the CIC estimator, we turn to the reduced third moment of the density field. The 
skewness $S_3$ is defined as 
\beq
\label{eqn:s3}
S_3\,\equiv\,{\displaystyle {\bar{\xi}}_3 \over \displaystyle {\bar{\xi}}_2^2}\,=\,{\displaystyle {\bar{\xi}}_3 \over \displaystyle \sigma^4}
\eeq

An estimate of $S_3$ can therefore be readily obtained on the basis of the corrected volume-averaged 3-point 
correlation function of the dark matter density field (see eqn.~\ref{cor-npoint-f} and eqn.~\ref{corrected-connected-moments}), 
\beq
{\widehat S_3}[CIC](R)\,=\,{\displaystyle {\bar{\xi}}_3 \over \displaystyle \sigma^4}\,=\,{\displaystyle k_3 \, \tilde{N}(R)^{-3} \over \displaystyle k_2^2 \, \tilde{N}(R)^{-4}}\,=\,{\displaystyle k_3 \over \displaystyle k_2^2}\,\tilde{N}(R)\,,
\label{eqn:s3cic}
\eeq
where ${\tilde N}(R)$ is the number of particles in spherical cells of radius $R$.

\begin{figure*}
\mbox{\hskip -0.5truecm\includegraphics[width=0.55\textwidth,angle=-90]{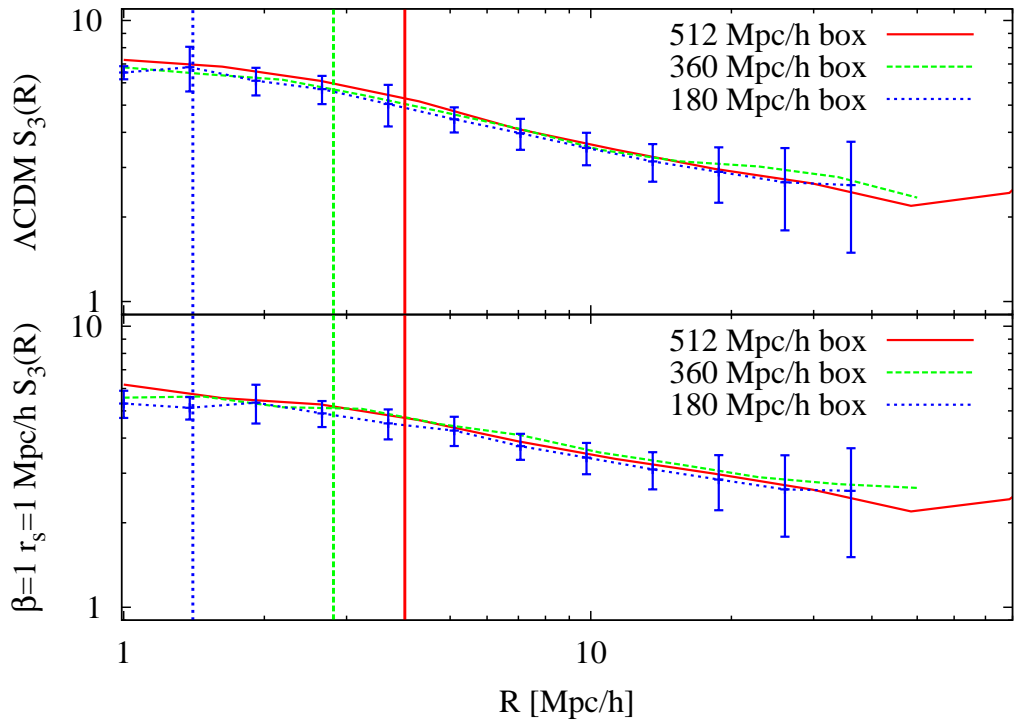}}
\caption{Comparison of the measured skewness $S_3(R)$ in ensembles with different box sizes. Top panel: results for 
pure $\lcdm$ model simulations. Bottom panel: results for the ensembles for a ReBEL model with scalar interaction 
parameters $\beta=1$ and $r_s=1\hmpc$. In each panel we plot the $S_3(R)$ relation for three different ensembles of 
the same model, one in a box with a width of $180\hmpc$ (blue dotted line), of $360\hmpc$ (green dashed line) and 
one of $512\hmpc$ (red solid line). The vertical lines show the corresponding values of the Nyquist scale 
$R_{Nyq}=2\pi/k_{Nyq}\sim2l$ for each of these simulation boxes. The error bars represent the $1\sigma$ errors in 
the 180LCDM (top panel) and 180B1RS1 (bottom panel) ensembles.}
\label{fig:s3_box_compare}
\end{figure*}
An alternative estimate of the skewness finds its origin in weakly nonlinear perturbation theory (PT, \cite{1980Peebles,
Juszkiewicz1993,skew_cur_pt,BCGS_book}). Juszkiewicz \textit{et al.}\cite{Juszkiewicz1993} showed that a good 
approximation for the skewness $S_3$ of the field, smoothed with the spherical top-hat window, is given by  
\beq
\label{eqn:s3_pt}
S_3\,=\,{34\over 7}\,+\,\gamma_1\,,
\eeq
where $\gamma_1$ is the logarithmic slope of the variance, defined as
\beq
\label{eqn:gamma1}
\gamma_1\,=\,-(n+3)\,=\,{d\log\sigma^2(R)\over d\log R}\,,
\eeq
where $n$ is the slope of the power spectrum at scale $R$. The term $34/7$ is a well-known result pertaining to 
the unsmoothed field (see \cite{1980Peebles}). For the estimate of the skewness ${\widehat S_3}[PT]$ based on this 
result, we use the estimate of the variance $\sigma^2(R)$ obtained via the integral over the non-linear 
power spectrum $P(k)$ for a tophat filter $W(kR)$, ie. from ${\widehat {\bar{\xi}}_2}[Pk](R)$ (eqn.~\ref{sigma-pk-pk}), 
\beq
\label{eqn:s3_pt_est}
{\widehat S_3}[PT](R)\,=\,{34\over 7}\,+\,{d\log {\widehat {\bar{\xi}}_2}[Pk](R)\over d\log R}\,.
\eeq

In fig.~\ref{fig:s3_test_scdm} we have compared the two estimates of the skewness $S_3$ for the SCDM 
simulations in the 1024SCDM ensemble, over a range of $1\hmpc < R < 80\hmpc$. The solid line represents the 
skewness measured directly on the basis of the counts in cells method (eqn.~\ref{eqn:s3cic}), while the dotted 
line is the perturbation theory prediction (eqn.~\ref{eqn:s3_pt_est}). The error bars, here shown for the 
CIC estimates, are the maximal standard deviation of the measurements for the simulations in the 1024SCDM 
ensemble (taking into account that at each different scale $R$ we use a \textit{different} number of 
sampling cells). 

Overall, we find that the two skewness estimators are in reasonable agreement with each other, in particular 
on linear and mildly nonlinear scales, $8\hmpc < R < 80\hmpc$, exactly as expected and reported by many other 
authors \cite{Juszkiewicz1993,cic_ana,npoint_omega_cdm}. 

\subsubsection{The box size test}
The effects of finite volume on the statistics of large scale structure have been extensively studied in 
several studies \cite{Colombi1994}. For most of the results presented in this paper, finite volume effects 
are rather unimportant. We focus mainly on a direct comparison between observables of the canonical 
$\lcdm$ model and those of the scalar-interacting dark matter ReBEL models. As long as any of the 
finite volume induced artefacts affects each of the cosmological models to a comparable extent, we need 
not worry about their influence on the results of our study. 

Nonetheless, there is one factor which needs to be investigated in some detail. The new physics of the 
dark sector scalar-interacting ReBEL models involves a new fundamental and intrinsic scale, the screening 
length $r_s$. It is a priori unclear in how far the relation between the length $L$ of the simulation box 
and the screening length $r_s$ of the ReBEL model will be of influence on the counts-in-cell measurement of 
various moments. 

To evaluate whether the finite box size has any impact on the measured values of $S_3$, we have run a set of 
simulations for two different models. One model is the $\Lambda$CDM model, whose gravitational force law is 
entirely scale-free, while the other model is a ReBEL model characterized by an intrinsic force scale. We chose 
a ReBEL model with strength factor $\beta=1$ and scale parameter $r_s=1 \hmpc$. Each of the two sets of 
simulations contain three ensembles of the same cosmological model. The first ensemble has a $180\hmpc$ 
simulation box, the second a $360\hmpc$ box and the third one a $512\hmpc$ box. 

In figure~\ref{fig:s3_box_compare} we follow the trend of the skewness $S_3(R)$ as a function of scale $R$, 
for each of the simulation ensembles. The top panel shows the results for the three sets of $\Lambda$CDM 
simulations, the 180LCDM simulations in a $180\hmpc$ box (dotted line), the 360LCDM simulations in a 
$360\hmpc$ box (dashed line) and the 512LCDM simulations in a $512\hmpc$ box (solid line). The same is  
repeated for the ReBEL model in the bottom panel, with the 180B1RS1 simulations in a $180\hmpc$ box (dotted line), 
the 360B1RS1 simulations in a $360\hmpc$ box (dashed line) and the 512LCDM simulations in a $512\hmpc$ box 
(solid line). In the figure we have also indicated the location of the Nyquist scale, $R_{Nyq}\equiv 2\pi/k_{Nyq}$, 
of each of the three simulation boxes. The three vertical lines mark their position. 

We find that in the $\lcdm$ case, the measured skewness in the simulation ensembles with different box size agree 
very well over the entire ranged we probed, from the largest measured scales $\sim 30\hmpc<R<80\hmpc$ down 
to the smallest scales of $1\hmpc<R<8\hmpc$. Interestingly, we also find a similar good agreement between 
the simulation ensembles of the ReBEL model. Moreover, we also find a surprisingly good agreement at 
scales where we expect two-body effects to start to dominate, below the Nyquist scale of the simulation.

Given the fact that the measured $S_3$ values remain consistent over such a wide range of scales and 
seems independent of the size of the simulation box size, we conclude that the effect of a different ratio 
$r_s/L$ of intrinsic force scale to box size has negligible, if any, effect on the measurement of 
statistical moments. 

\subsubsection{Transients}
The \verb#PMcode# that we use to generate the initial conditions is based on the Zeldovich Approximation (ZA) method\cite{ZA}. 
It is well known that the Zeldovich approximation introduces an artificial level of skewness and additional higher order 
hierarchy moments into the density field \cite{transients1,transients2}. A sufficient number of simulation time-steps is 
required for the true particle dynamics to take over and to relax these transient artifacts. An alternative approach 
is to resort to second order Lagrangian perturbation theory schemes for setting up the initial conditions of 
simulations \cite{scoccimarro98,transients1,transients2,jenkins10}.

Because of the above, the initial redshift of a cosmological simulation is an important factor in determining the statistical 
reliability of the cosmological numerical experiment. In general, for the purpose of comparing density fields and cumulants 
in different models we need to be less concerned about the net amplitude of the transients as they will have the same 
magnitude in all models. 

Nonetheless, there is an additional factor that depends on the initial redshift and which only affects the ReBEL 
models. The intrinsic scalar force of these models should be able to act as long as possible, in order to account 
for an optimal representation of their impact on the dark matter density field. If the ReBEL simulations are evolved 
too far by means of the Zeldovich approximation and their dynamical evolution started too late, the deviation of the 
ReBEL dark matter density field from the one in the $\Lambda$CDM simulations will diminish. 

\begin{figure}[h]
\mbox{\hskip -0.5truecm\includegraphics[width=67mm,angle=-90]{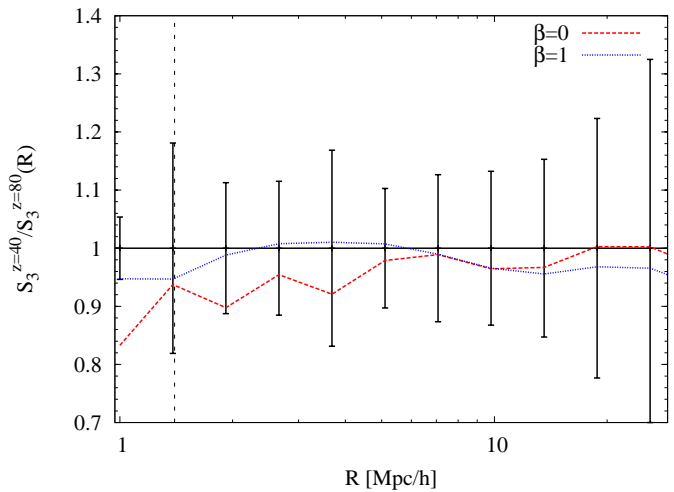}}
\caption{ Test of transients effects for $180\hmpc$ ensembles. $S_3(z_i=40)/S_3(z_i=80)$: 
ratio of skewness $S_3(z_i=40)$ measured for the simulation ensemble started at redshift $z_i=40$ to the skewness 
of the simulations started at redshift $z_i=80$. The ratio $S_3(z_i=40)/S_3(z_i=80)$ is plotted as a function 
of scale $R$. The solid black horizontal line represents the unity line for which $S_3(z_i=40)=S_3(z_i=80)$. The 
dashed vertical line marks the Nyquist scale for the simulations in a $180\hmpc$ box, at $2\pi/k_{Nyq}=1.38\hmpc$. 
The ratio $S_3(z_i=40)/S_3(z_i=80)$ is plotted for two different situations. Red dashed line: skewness ratio  for the two 
ensembles of $\Lambda$CDM simulations, 180LCDM and 180LCDMZ80. Blue dotted line: skewness ratio for the two 
ReBEL simulation ensembles with $\beta=1$ and $r_s=1\hmpc$, 180B1RS1 and 180B1RS1Z80. The error bars represent 
the $1\sigma$ values determined for the 180LCDM ensemble. The errors in the other ensembles have a similar 
magnitude.}
\label{fig:trans}
\end{figure}

In order to quantify the possible effects of the transients, we have performed a series of auxiliary simulation 
ensembles. These contain $256^3$ DM particles placed in boxes of the box width $180\hmpc$ and have 10 times better force resolution. 
There are two ensembles, one  for the $\lcdm$ model, 180LCDMZ80, and one for the ReBEL model with $\beta = 1$ and $r_s = 1\hmpc$, 
180B1RS1Z80. We will compare them with our main ensembles for the same models, 180LCDM and 180B1RS1. Therefore the ensembles 
of each model will differ only in the force resolution and the redshift at which the N-body calculation 
is started, one at $z_i=80$ and the other at $z_i=40$. 

The results for the direct comparison of the skewness $S_3$ in the $z_i=80$ models and the $z_i=40$ models, 
in terms of their ratio $S^{z_i=40}_3/S^{z_i=80}_3$, are plotted in figure~\ref{fig:trans}. The blue dotted line 
represents the ratio for the ReBEL ensemble, the red dashed line for the $\Lambda$CDM ensemble (for which 
$\beta=0$). For reference, the black horizontal solid line indicates the unity ratio $S^{z_i=40}_3/S^{z_i=80}_3=1$, 
while the vertical line marks the Nyquist scale $2\pi/k_{Nyq} \cong 1.4\hmpc$ for these simulations. The 
error bars mark $1\sigma$ errors in the 180LCDM ensemble, with errors in the other three ensembles being 
of the same order. 

We note that the visible transients effects are, if real, very small. The skewness ratio curves lie 
very close to the unity line $S^{z_i=40}_3/S^{z_i=80}_3=1$. Their deviations from unity are smaller than 
the $1\sigma$ errors, with discrepancies not exceeding the $10\%$ level. On the basis of this we may 
conclude that the redshift of the initial conditions of our main ensembles, at $z_i=40$, is sufficiently 
high to assure that any effects of possible transient are negligible for our analysis. 

\subsubsection{Resolution}

\begin{figure}[h]
\mbox{\hskip -0.5truecm\includegraphics[width=67mm,angle=-90]{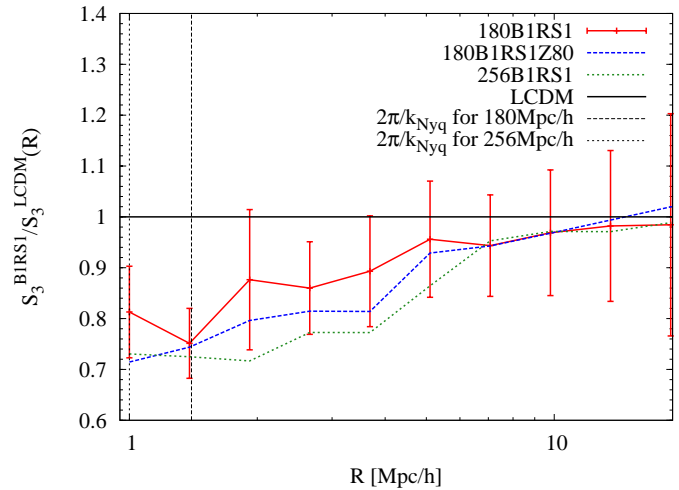}}
\caption{(color-on-line) Test of resolution effects on computed skewness $S_3$. Plotted 
are the ratios $S^{B1RS1}_3/S^{LCDM}_3$ of the skewness obtained in the ReBEL model with $\beta=1$ and 
$r_s=1\hmpc$ to the skewness measured in the LCDM model, for three simulations with different force and 
mass resolutions. The lines depict relevant ratios for lower resolution run 180B1RS1 (solid line), 
high force resolution run 180B1RS1Z80 (dashed line) and high mass and force resolution run 256B1RS1HR 
(dotted line). The vertical dashed line marks Nyquist scale for runs with $256^3$ particles, while the 
vertical dotted line depicts the same scale for $512^3$ particles simulations. The error bars represent 
the $1\sigma$ values determined for the 180B1RS1 ensemble.}
\label{fig:resolution}
\end{figure}

The last important effect we must check is the impact of the mass and force resolution used in our simulations on the 
measured quantities. The mass resolution is related to the mean inter-particle separation, while the force resolution 
corresponds to the scale at which the force prescription of the simulation code exactly recovers the intended 
Newtonian - or ReBEL - force.

To investigate the impact of these resolution factors on the measurement skewness we use the high force resolution ensembles 
180LCDMZ80 and 180B1RS1Z80, as well as two single high mass and force resolution runs, 256LCDHR and 256B1RS1HR 
(see table~\ref{tab:sim_params}). For these two simulations we use bootstrap resampling to obtain averages of 
mean and variance of the measured moments. This is accomplished as follows. We randomly cast a large number of 
spherical cells over the entire simulation volume. This ranges from $2\times10^8$ cells with $R=1\hmpc$ to 
$2\times 10^6$ cells for $R=30\hmpc$. Ten sets of measurements were constructed, each consisting of a random 
subset of $10\%$ of the casted spheres.

We may assess the resolution effects on the basis of the plot in figure \ref{fig:resolution}. It depicts the 
ratio of the skewness in three different ReBEL ensembles to that of the skewness in the LCDM model, 
$S^{B1RS1}_3/S^{LCDM}_3$. Each of the three ReBEL models have the same ReBEL parameters, $\beta=1$ and 
$r_s=1\hmpc$, but differ in resolution. The lower resolution run is 180B1RS1 (solid line), the high 
force resolution run is 180B1RS1Z80 (dashed line), while the high mass plus high force resolution run 
is that of 256B1RS1HR (dotted line). The error-bars marking the skewness ratio of the 180B1RS1 
run are the $1\sigma$ errors for 180B1RS1 ensemble.

Even though we find that the simulations with a higher resolution show a systematically higher signal 
level at scales $R<7\hmpc$, this effect is entirely contained within - or at best marginally above - the 
$1\sigma$ errors of the 180LCDM ensemble. We may therefore conclude that an increase in the force and/or mass 
resolution of the simulation does not yield a significant improvement of the signal level. This reassures 
us that the simulation ensembles used in our main study yield good and reliable estimates of the 
quantities which we study.

\subsection{CIC test summary}
In all, we may conclude from the various tests of the Count-in-Cell method that it is perfectly suited for 
studying the impact impact of long-range scalar interactions on the higher-order correlation statistics of 
the dark matter density field.

\begin{figure*}
\mbox{\hskip -0.62cm\includegraphics[width=0.54\textwidth,angle=-90]{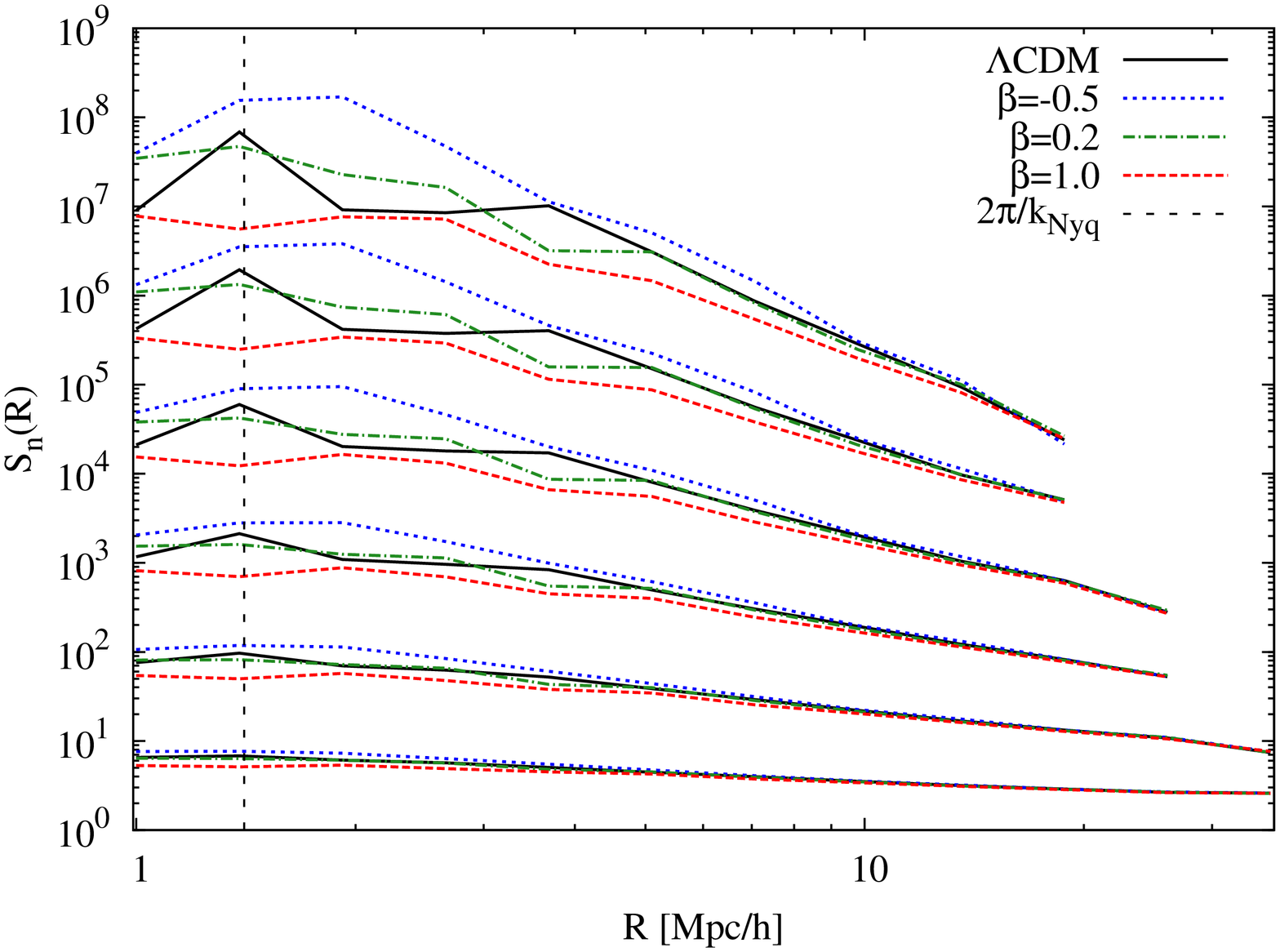}}
\caption{Hierarchical amplitude $S_n(R)$ as a function of scale $R$. Plotted are $S_n$, for $n=3$ to 
$n=8$, for four different simulation ensembles: the canonical $\lcdm$ model 180LCDM simulations (solid black line), the 
$\beta=-0.5$ and $r_s=1 \hmpc$ ReBEL model simulations 180B-05RS1 (blue dotted line), the 
$\beta=0.2$ and $r_s=1 \hmpc$ ReBEL model simulations 180B02RS1 (green dot-dashed line) and the 
$\beta=1.0$ and $r_s=1 \hmpc$ ReBEL model simulations 180B1RS1 (red dashed line). The $S_3$ curves 
have the lowest amplitude, with the amplitude of the $S_n$ curves systematically increasing as a function 
of $n$. The thin vertical dashed line marks the Nyquist scale $R_{Nyq}\approx 1.4\hmpc$.\label{fig:180_moments}}
\includegraphics[width=0.53\textwidth,angle=-90]{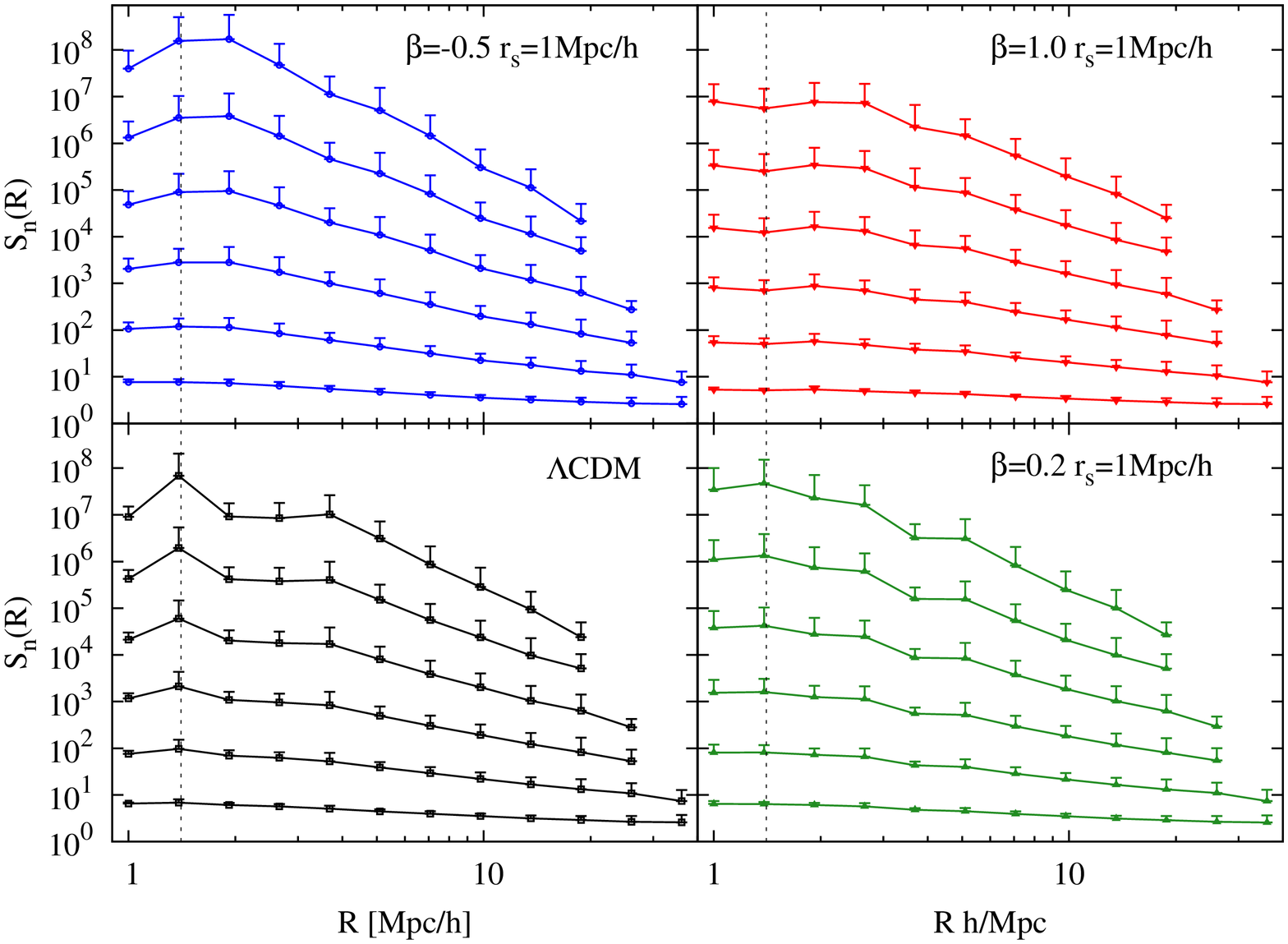}
\caption{Hierarchical amplitudes $S_n(R)$ as a function of scale $R$. Equivalent to fig.~\ref{fig:180_moments}, 
the $S_n$ are plotted in a separate panel for each cosmological model. Top left: $\beta=-0.5$ and $r_s=1 \hmpc$ ReBEL 
model simulations 180B-05RS1; top right: $\beta=1.0$ and $r_s=1 \hmpc$ ReBEL model simulations 180B1RS1; 
bottom left: the canonical $\lcdm$ model 180LCDM simulations; bottom right: $\beta=0.2$ and $r_s=1 \hmpc$ ReBEL model 
simulations 180B02RS1. The $S_3$ curves have the lowest amplitude, with the amplitude of the $S_n$ curves systematically 
increasing as a function of $n$. The thin vertical dashed line marks the Nyquist scale $R_{Nyq}\approx 1.4\hmpc$.\label{fig:180_moments2}}
\end{figure*}

\section{Moment Analysis of N-body ensembles}
\label{sec-results}
Having ascertained ourselves of the reliability of the CIC machinery, we will present and discuss the results 
of the correlation function analysis of our N-body experiments. The intention of this study is the identification 
of discriminative differences between the canonical $\lcdm$ cosmology and a range of scalar interaction ReBEL models. 

\begin{figure*}
\includegraphics[width=1.15\textwidth,angle=-90]{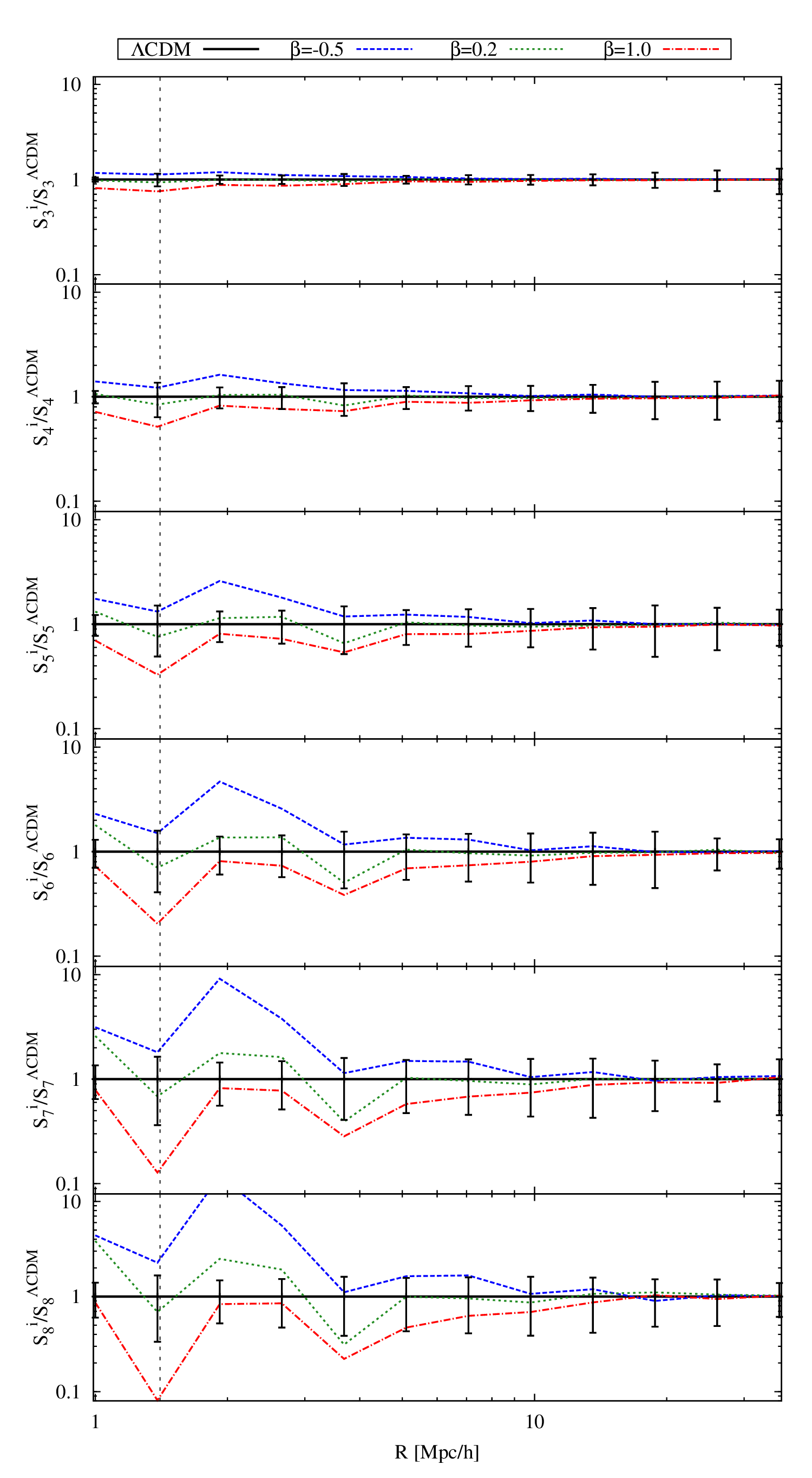}
\caption{Hierarchical amplitude $S_n(R)$ as a function of scale $R$: ratio $S_n(R)/S_n^{\lcdm}(R)$ of 
the hierarchy amplitudes in any of the simulation ensembles to the hierarchy amplitudes in the concordance 
$\Lambda$CDM cosmology simulations. From top to bottom panel: $S_3/S_3^{\lcdm}$ to $S_8/S_8^{\lcdm}$. In each 
panel we plot the curves for four different simulations ensembles: the canonical $\lcdm$ model 180LCDM simulations, 
by definition equal to unity (solid black line), the $\beta=-0.5$ and $r_s=1 \hmpc$ ReBEL model simulations 180B-05RS1 
(blue dotted line), the $\beta=0.2$ and $r_s=1 \hmpc$ ReBEL model simulations 180B02RS1 (green dot-dashed line) and the 
$\beta=1.0$ and $r_s=1 \hmpc$ ReBEL model simulations 180B1RS1 (red dashed line).  The $S_3$ curves 
 have the lowest amplitude, with the amplitude of the $S_n$ curves systematically increasing as a function 
 of $n$. The thin vertical dashed line marks the Nyquist scale $R_{Nyq}\approx 1.4\hmpc$. 
The error-bars correspond to $1\sigma$ scatter of the 180LCDM ensemble.
\label{fig:180_moments_diff}}
\end{figure*}

We address two aspects of the resulting dark matter distributions. The first concerns a complete census of the 
hierarchy amplitudes $S_n$, from $n=3$ to $n=8$ for a set of three different ReBEL model simulations and a similar  
ensemble of $\Lambda$CDM simulations. In addition, in order to assess the redshift evolution of these statistical 
measures, we focus on the redshift dependence of the skewness and kurtosis. 

\subsection{Hierarchy amplitudes}
The hierarchy amplitudes $S_n(R)$ of order $n$ are conventionally defined as,
\beq
\label{eqn:corr_amp}
S_n(R)\,=\,{\displaystyle \bar{\xi_n}\over \displaystyle \bar{\xi_2}^{n-1}}\,=\,\bar{\xi_n}\,\sigma^{-2(n-1)}\;,
\eeq
with the volume-averaged correlation functions $\bar{\xi_n}(R)$ and variance $\sigma^2(R)$ implicitly depending 
on the scale $R$. 

\subsubsection{General Trends}
In figure~\ref{fig:180_moments} and \ref{fig:180_moments2} we plot the measured $S_n$'s, from $n=3$ up to $n=8$, for 
all simulation ensembles with boxwidth $180\hmpc$ (see table~\ref{tab:sim_params}). These two figures represent the 
key result of this study. 

The volume-averaged N-point correlation functions ${\bar \xi_n}$ have been computed by means of the CIC method, following the 
description in section~\ref{sec-method}. The simulations for which the hierarchy amplitudes have been computed are the 180LCDM 
set of $\Lambda$CDM simulations and the 180B-05RS1, 180B02RS1 and 180B1RS1 simulations of the ReBEL models with scalar interaction 
scale parameter $r_s=1\hmpc$ and strength parameter $\beta=-0.5$, $\beta=0.2$ and $\beta=1.0$. The $\beta = -0.5$ case, 
whose physical effect is that of a repulsive scalar ReBEL force, does not have a real physical motivation. It is mainly 
included for reference, in order to outline the impact of the $\beta$ strength parameter on the final nonlinear density field. 
For all model simulations we have calculated the hierarchy amplitudes $S_n$ at 12 logarithmically spaced scales within the 
range of $1\hmpc < R < 36.09\hmpc$. The exact values of these 12 scale values $R$ are listed in table~\ref{tab:s3}. 

Figures~\ref{fig:180_moments} and \ref{fig:180_moments2} plot the hierarchy amplitudes $S_n$ as a function of scale $R$. 
The two figures are complementary: in figure~\ref{fig:180_moments2} the $S_n$ are shown separately for each of the 
cosmological models, while figure~\ref{fig:180_moments} superimposes the curves for each of the models in order to highlight 
their differences. In addition, to provide an impression of the relative differences between hierarchy amplitudes in 
each of the cosmological models, figure~\ref{fig:180_moments_diff} plots the ratio between the $S_n(R)$ between each ReBEL model 
and the concordance $\lcdm$ models. In figure~\ref{fig:180_moments}, each cosmological model is indicated by a different line 
types. The canonical $\lcdm$ model is indicated by the black solid line, the ReBEL model with $\beta = -0.5$ and $r_s=1\hmpc$ 
by the blue dotted line, the ReBEL model with $\beta = 0.2$ and $r_s=1\hmpc$ by the green dot-dashed line and the ReBEL 
model with $\beta = 1.0$, $r_s=1\hmpc$ by the red dashed line. Figure~\ref{fig:180_moments2} also includes the error bars 
of the measured $S_n$ values, restricted to their upper half for purposes of clarity. For reference, we have listed 
the values of the standard deviation for the skewness $S_3$ and kurtosis $S_4$ in tables~\ref{tab:s3} and \ref{tab:s4}. 
The thin dashed vertical lines in figures~\ref{fig:180_moments} and \ref{fig:180_moments2} mark the Nyquist scale 
$R_{Nyq}=2\pi/k_{Nyq} \approx 1.4 \hmpc$ for the $\Lambda$CDM and ReBEL simulations, which for these realizations is 
double the mean inter-particle separation $2l$. We consider the computed quantities on scales below the Nyquist scale 
as unreliable, and exclude them from further analysis in this study. 

There are some clear trends in the behaviour of the $S_n$ hierarchy. At large scales, $R>10\hmpc$, all cosmologies 
agree on the $S_n$. This is straightforward to understand because at these large scales the ReBEL models are 
practically equivalent to the $\Lambda$CDM cosmology. The differences between the models become distinct at scales 
$R\leq 10\hmpc$, where the effect of the scalar ReBEL force kicks in. We discern a systematic trend, with all $S_n$ 
consistently higher than the $\lcdm$ values for the ReBEL model with $\beta=-0.5$, consistently lower than the 
$\lcdm$ values for the ReBEL model with $\beta=1.0$ and the values for the ReBEL model with $\beta=0.2$ straddling 
tightly around the $\lcdm$ values. We also notice that the differences between the models increase systematically 
as a function of order $n$ (see fig.~\ref{fig:180_moments_diff}). This may be easily understood from the higher sensitivity 
of the higher moments to the changing shape of the density probability function, and hence to the changes in the dark matter 
density distribution. 

\begin{table}[h!]
\caption{\label{tab:s3} Measured values of the $S_3$ hierarchy amplitude at redshift $z=0$, for four 
simulation ensembles in different cosmologies. The models are the $\lcdm model$ 180LCDM simulation, the 
$\beta=-0.5$ and $r_s=1 \hmpc$ ReBEL model simulations 180B-05RS1, the $\beta=0.2$ and $r_s=1 \hmpc$ ReBEL 
model simulations 180B02RS1 and the $\beta=1.0$ and $r_s=1 \hmpc$ ReBEL model simulations 180B1RS1.}
\medskip
\begin{tabular}{cllll}
\hline
&&&& \\
R & 180LCDM & 180B-05RS1 & 180B02RS1 & 180B1RS1 \\
($\hmpc$) &&&&\\
&&&& \\
\hline 
&&&& \\
01.00 & $6.53\pm0.35$ & $7.65\pm1.08$ & $6.40\pm1.00$ & $5.30\pm0.59$\\
01.38 & $6.82\pm1.23$ & $7.69\pm1.23$ & $6.35\pm0.74$ & $5.13\pm0.47$\\
01.92 & $6.10\pm0.69$ & $7.29\pm1.45$ & $6.09\pm0.70$ & $5.35\pm0.84$\\
02.66 & $5.70\pm0.66$ & $6.35\pm1.37$ & $5.66\pm1.00$ & $4.90\pm0.52$\\
03.68 & $5.05\pm0.85$ & $5.48\pm0.87$ & $4.80\pm0.39$ & $4.51\pm0.55$\\
05.10 & $4.45\pm0.46$ & $4.72\pm0.81$ & $4.49\pm0.73$ & $4.26\pm0.51$\\
07.07 & $3.97\pm0.50$ & $4.07\pm0.61$ & $3.90\pm0.51$ & $3.74\pm0.40$\\
09.80 & $3.52\pm0.46$ & $3.56\pm0.50$ & $3.50\pm0.47$ & $3.41\pm0.43$\\
13.57 & $3.15\pm0.48$ & $3.21\pm0.54$ & $3.12\pm0.47$ & $3.10\pm0.47$\\
18.80 & $2.89\pm0.65$ & $2.90\pm0.66$ & $2.87\pm0.65$ & $2.85\pm0.63$\\
26.05 & $2.66\pm0.86$ & $2.68\pm0.87$ & $2.66\pm0.87$ & $2.63\pm0.84$\\
36.09 & $2.60\pm1.10$ & $2.60\pm1.10$ & $2.57\pm1.08$ & $2.60\pm1.09$\\
&&&& \\
\hline
\end{tabular}
\bigskip
\caption{\label{tab:s4} Measured values of the $S_4$ hierarchy amplitude at redshift $z=0$, for four 
simulation ensembles in different cosmologies. The models are the $\lcdm model$ 180LCDM simulation, the 
$\beta=-0.5$ and $r_s=1 \hmpc$ ReBEL model simulations 180B-05RS1, the $\beta=0.2$ and $r_s=1 \hmpc$ ReBEL 
model simulations 180B02RS1 and the $\beta=1.0$ and $r_s=1 \hmpc$ ReBEL model simulations 180B1RS1.}
\medskip
\begin{tabular}{cllll}
\hline
&&&& \\
R & 180LCDM & 180B-05RS1 & 180B02RS1 & 180B1RS1\\
($\hmpc$) &&&&\\
&&&& \\
\hline
&&&& \\
01.00 & $75\pm12$ & $106\pm40$ & $81\pm39$ & $54\pm19$\\
01.38 & $97\pm55$ & $118\pm58$ & $82\pm35$ & $50\pm16$\\
01.92 & $70\pm20$ & $113\pm68$ & $72\pm26$ & $57\pm25$\\
02.66 & $63\pm20$ & $84\pm54$ & $66\pm33$ & $48\pm16$\\
03.68 & $52\pm28$ & $61\pm27$ & $43\pm9$ & $38\pm13$\\
05.10 & $39\pm12$ & $44\pm23$ & $40\pm19$ & $35\pm12$\\
07.07 & $29\pm10$ & $31\pm14$ & $28\pm11$ & $26\pm8$\\
09.80 & $22\pm8$ & $22\pm9$ & $22\pm8$ & $20\pm7$\\
13.57 & $17\pm7$ & $18\pm8$ & $16\pm7$ & $16\pm7$\\
18.80 & $13\pm8$ & $13\pm9$ & $13\pm8$ & $13\pm8$\\
26.05 & $11\pm7$ & $11\pm7$ & $11\pm7$ & $11\pm7$\\
36.09 & $7\pm5$ & $8\pm5$ & $7\pm5$ & $8\pm5$\\
&&&& \\
\hline
\end{tabular}
\end{table}

\subsubsection{Skewness and Kurtosis}
In the observational reality, beset by various sources of noise, it may be cumbersome to get reliable 
estimates of higher order moments. On the other hand, we may expect reasonably accurate estimates of the 
third and fourth order moments, the skewness and kurtosis. The question is whether the presence or absence 
of ReBEL scalar forces may be deduced from the behaviour of these moments. To evaluate the discriminatory powers 
of $S_3$ and $S_4$ we list the measured values of these hierarchy amplitudes in tables~\ref{tab:s3} and 
~\ref{tab:s4}.  

Assessing the data presented in these tables reveals that $S_3$ and $S_4$ values converge to within 
$1\sigma$ around the $\lcdm$ values for scales larger than $R=9.8 \hmpc$. As we turn towards smaller 
scales $R$, the ReBEL model values for the skewness and kurtosis display an increasingly large difference 
with respect to the $\lcdm$ value. In other words, at these small (mildly) nonlinear scales we observe 
a direct imprint of the scalar forces on the density field moments. 

At scales comparable to the screening length, ReBEL models with a positive strength parameter $\beta$ have a lower 
skewness and kurtosis value than those for the canonical $\lcdm$ model. The difference is smaller 
for ReBEL models with a lower $\beta$, and turns into a higher value as $\beta$ turns negative. Seen as a 
function of scale, the difference decreases towards larger scales $R$. 

For the $\beta=1.0$ 180B1RS1 simulations the value of $S_3$ at $R \sim 1.4\hmpc$ is $\approx 25\%$ lower than 
the value for the $\lcdm$ model, while the discrepancy is in only the order of $\approx 10\%$ at $R=3.68\hmpc$ 
and has dropped towards $\leq 5\%$ for $R\geq 7\hmpc$. The differences are more prominent in the case of the 
kurtosis $S_4$. For $R\sim 1.4\hmpc$ the value of $S_4$ is smaller than the $\lcdm$ value by no less than 
$\approx 48\%$, decreasing towards $\approx 27\%$ at $R=3.68$ and to less than $10\%$ at $R\geq 7\hmpc$.

The differences between the cosmological models are therefore less substantial for the skewness $S_3$ than for 
the kurtosis $S_4$. On the condition that it is possible to obtain reliable estimates for $S_4$ in the 
observational reality, this leads us to the conclusion that the kurtosis may be better suited as tracer of 
ReBEL signatures in the density field. More detailed studies and simulations, including baryons and mock galaxy 
samples, will be necessary to make a final choice for the optimal marker of ReBEL cosmology in observational 
catalogues. 

\begin{figure}
\mbox{\hskip -0.5truecm\includegraphics[width=67mm,angle=-90]{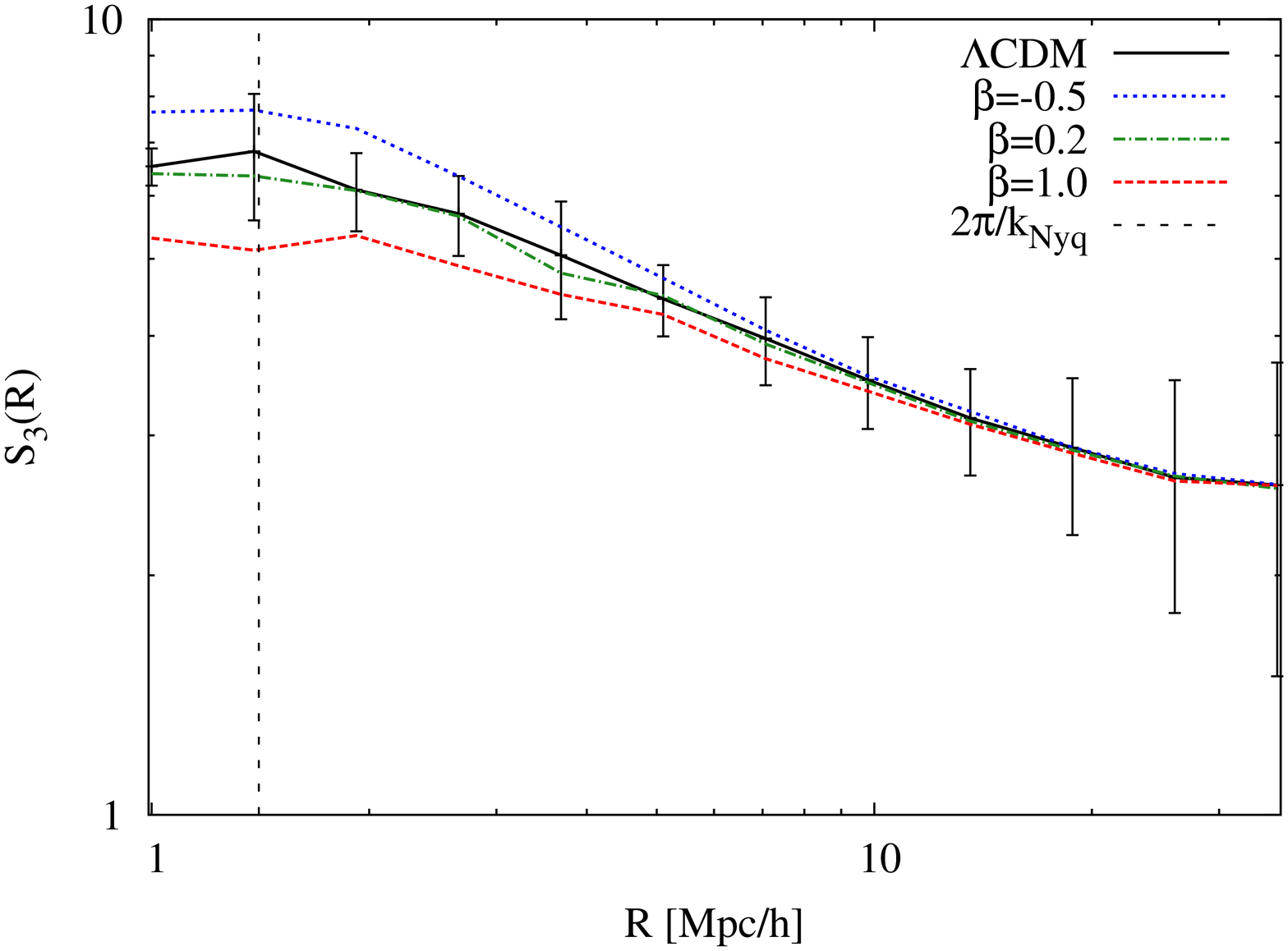}}
\mbox{\hskip -0.5truecm\includegraphics[width=67mm,angle=-90]{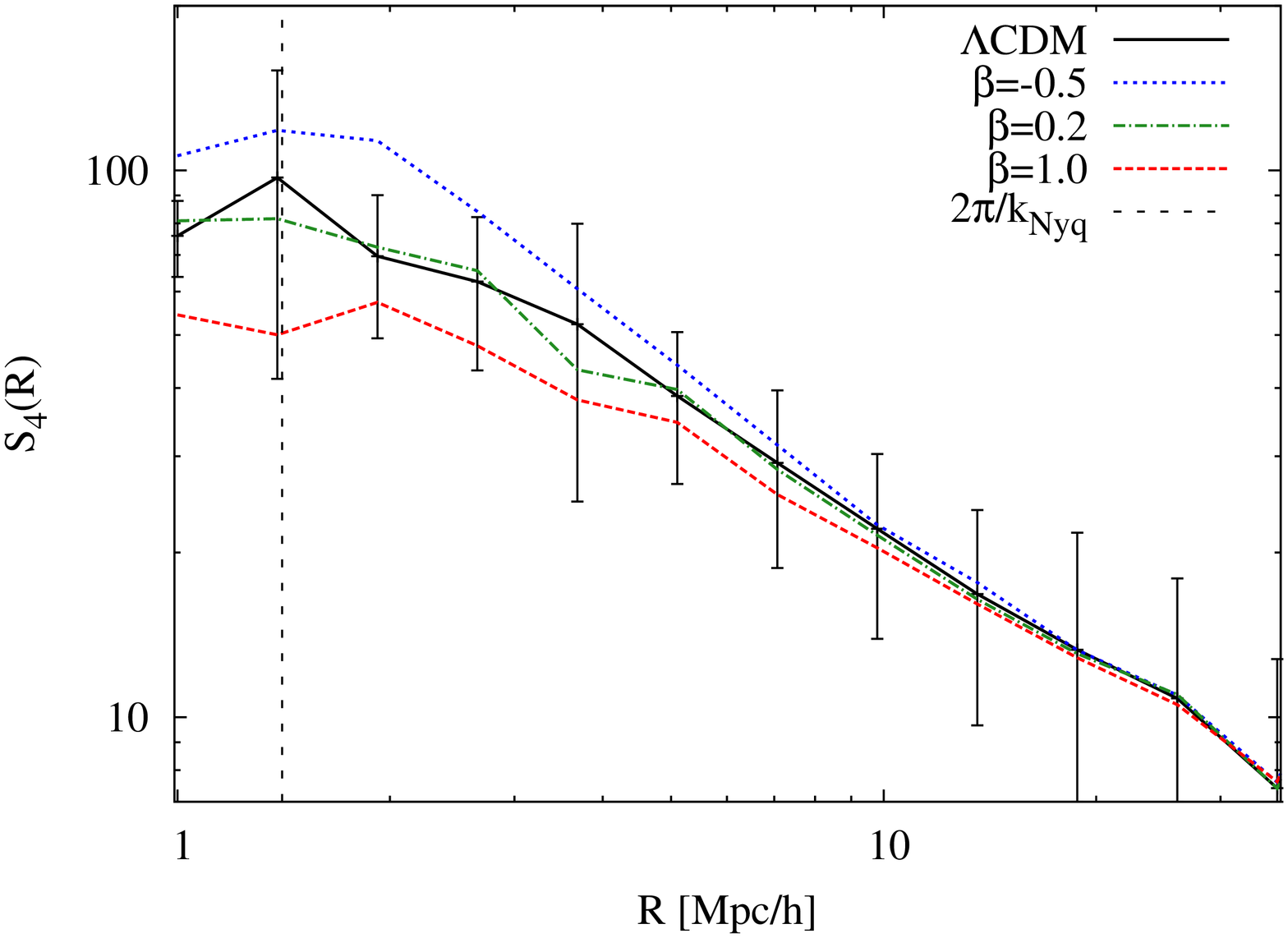}}
\caption{Skewness $S_3(R)$ and kurtosis $S_4(R)$ as a function of scale $R$. Plotted are $S_3(R)$ and $S_4(R)$ 
in four $180\hmpc$ box simulation ensembles: the canonical $\lcdm$ model 180LCDM simulations (solid black line), the 
$\beta=-0.5$ and $r_s=1 \hmpc$ ReBEL model simulations 180B-05RS1 (blue dotted line), the 
$\beta=0.2$ and $r_s=1 \hmpc$ ReBEL model simulations 180B02RS1 (green dot-dashed line) and the 
$\beta=1.0$ and $r_s=1 \hmpc$ ReBEL model simulations 180B1RS1 (red dashed line). The thin vertical dashed 
line marks the Nyquist scale $R_{Nyq}\approx 1.4\hmpc$. The error-bars correspond to the $1\sigma$ scatter 
of the 180LCDM ensemble.}
\end{figure}

\begin{table}[h!]
\caption{\label{tab:s3s4_z} Values of the deviations $\Delta S_3$ and $\Delta S_4$ of the skewness and kurtosis, 
at a scale of $R\sim 1.4\hmpc$, measured for the ReBEL models from those of the canonical $\lcdm$ model. For 
the definition of $\Delta S_3$ and $\Delta S_4$ see Eqn.~(\ref{eqn:dev_def}). First column: redshift $z$. 
Three additional columns: values for $S_3$ (top table) and $S_4$ (bottom table) for three $180\hmpc$ ReBEL 
ensemble simulations, 180B-05RS1, 180B02RS1 and 180B1RS1.}
\begin{tabular}{cccccccc}
&&&&&&&\\
\hline
\hline
&&&&&&&\\
\multicolumn{8}{c}{$\Delta S_3$}\\
&&&&&&&\\
\hline
&&&&&&&\\
z &&& 180B-05RS1 && 180B02RS1 && 180B1RS1\\
&&&&&&&\\
\hline 
&&&&&&&\\
0.0 &&& 0.126 && -0.066 && -0.247\\
0.5 &&& 0.295 && -0.038 && -0.245\\ 
1.0 &&& 0.270 && -0.103 && -0.328\\
2.0 &&& 0.220 && -0.100 && -0.305\\
5.0 &&& 0.020 && -0.014 && -0.120\\
&&&&&&&\\
\hline
\hline
&&&&&&&\\
&&&&&&&\\
\hline
\hline
&&&&&&&\\
\multicolumn{8}{c}{$\Delta S_4$}\\
&&&&&&&\\
\hline
&&&&&&&\\
z &&& 180B-05RS1 && 180B02RS1 && 180B1RS1\\
&&&\\
\hline
&&&&&&&\\
0.0 &&& 0.216 && -0.154 && -0.484\\
0.5 &&& 0.698 && -0.075 && -0.471\\
1.0 &&& 0.560 && -0.219 && -0.603\\
2.0 &&& 1.050 && -0.242 && -0.550\\
5.0 &&& 0.600 && -0.045 && -0.288\\
&&&&&&&\\
\hline
\hline
\end{tabular}
\end{table}

\begin{figure*}
\mbox{\hskip -0.4truecm\includegraphics[width=0.775\textwidth,angle=-90]{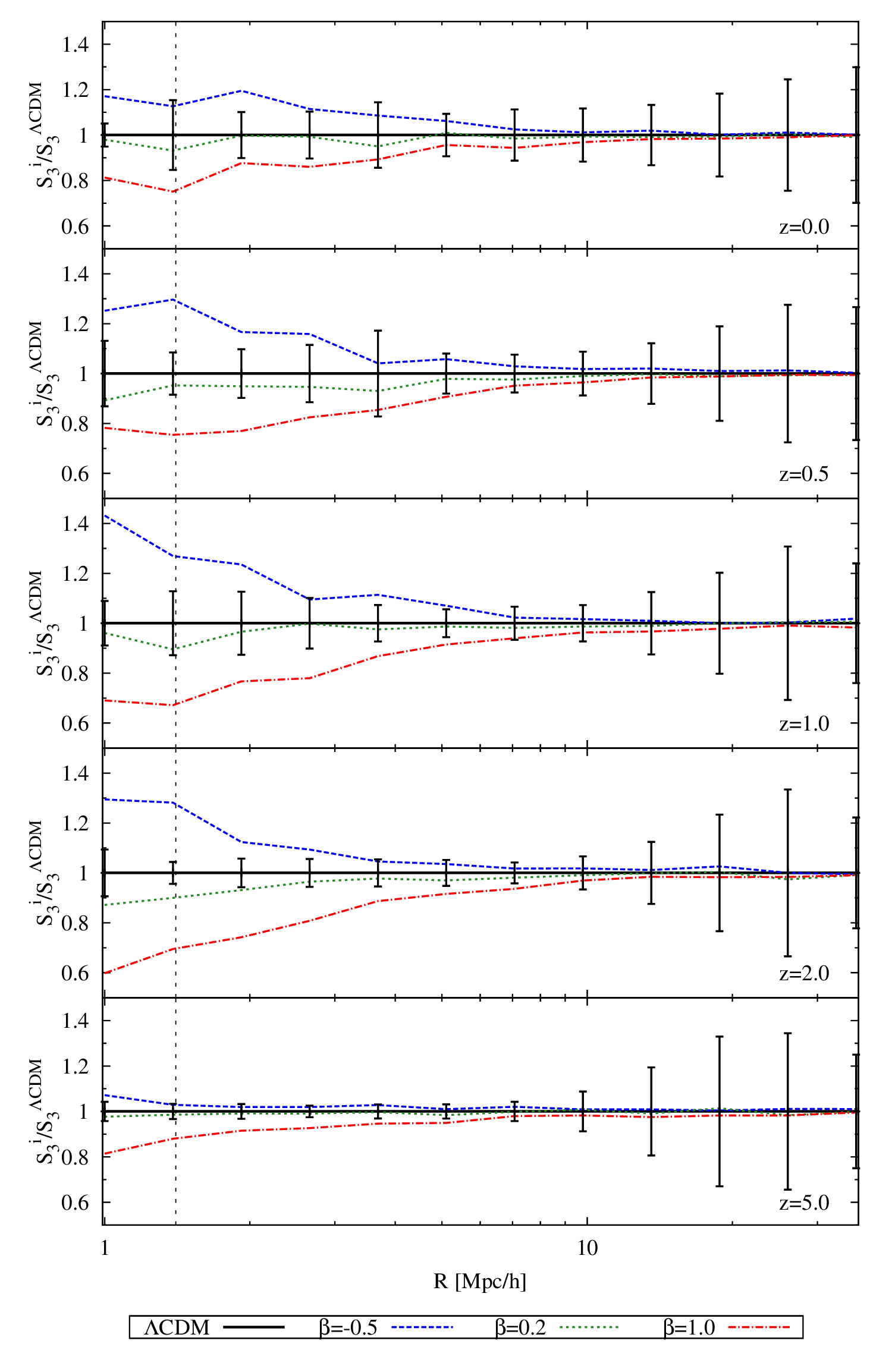}}
\includegraphics[width=0.775\textwidth,angle=-90]{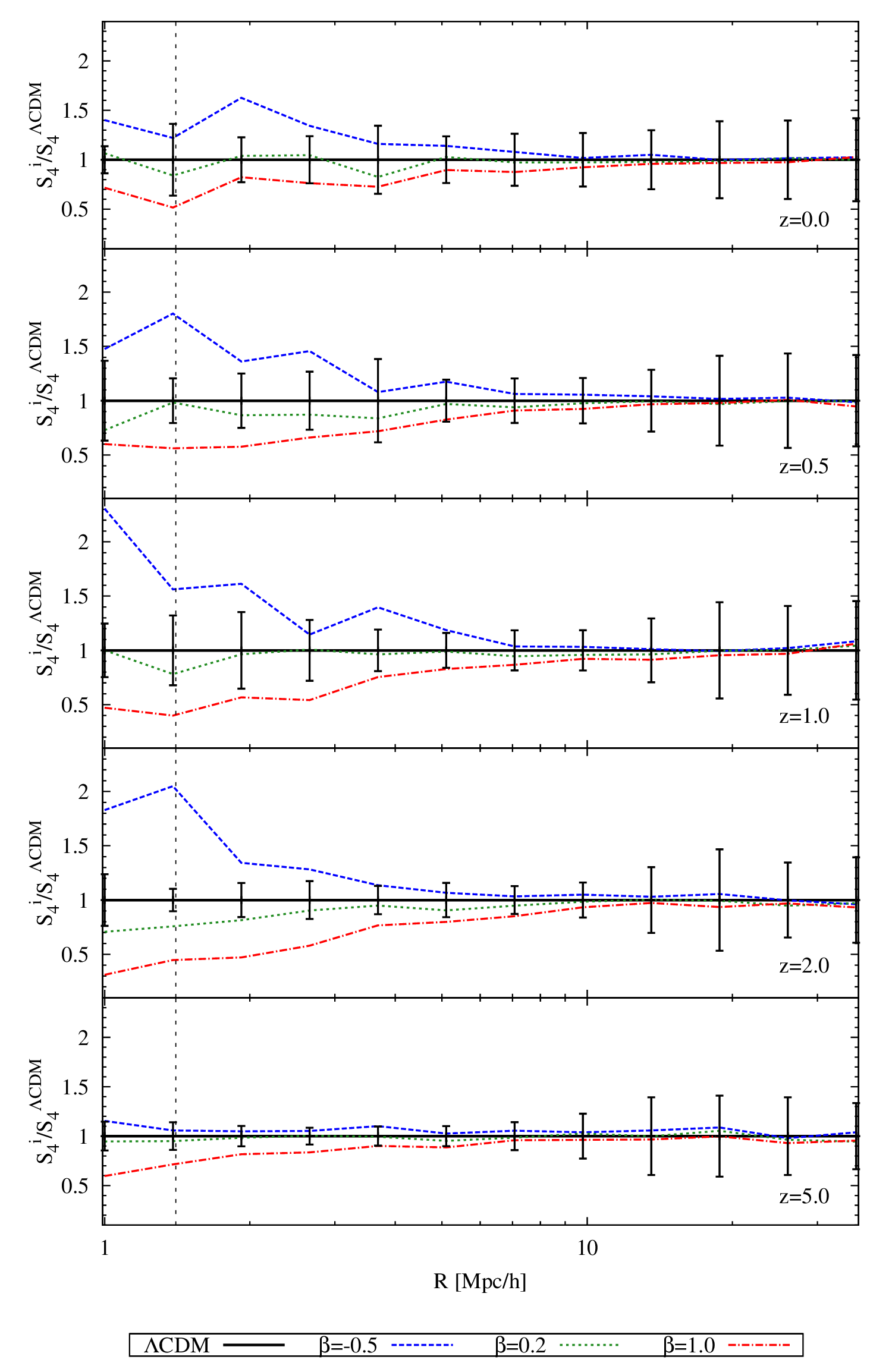}
\caption{Left column: redshift evolution of the skewness ratios $S^{ReBEL}_3/S^{\lcdm}_3$ for different ReBEL models. 
The lines mark the ensembles 180LCDM (solid line), 180B-05RS1(dashed line), 180B02RS1 (dotted line) and 
180B1RS1(dotted-dashed line). Panels shows redshifts for $z=0$ (the top panel) to $z=5$ (the bottom panel). 
Vertical dashed line marks the Nyquist scale $\sim 1.4\hmpc$. Righthand panel: identical set of redshift panels 
for the kurtosis ratio $S^{ReBEL}_4/S^{\lcdm}_4$. The error-bars correspond to the $1\sigma$ scatter of 
the 180LCDM ensemble.}
\label{fig:s3_s4_z}
\end{figure*}

\subsection{Redshift evolution}
In our previous study \cite{LRSI1} we found that the amplitude of the deviation of the two-point correlation 
function $\bar{\xi}$ of the ReBEL model to that of the $\lcdm$ model changes with redshift. This suggests a 
similar evolution of higher order moments like $S_3$ and $S_4$, prodding us to assess the redshift evolution 
of skewness and kurtosis. 

To this end, we study the archive of five snapshots -- at redshifts $z=5., 2., 1., 0.5$ and $z=0$ -- which 
for each simulation in the four $180\hmpc$ ensembles were saved: 180LCDM, 180B-05RS1, 180B02RS1, and 180B1RS1. 

The redshift evolution of the skewness and kurtosis in the four simulation ensembles can be followed in 
figure~\ref{fig:s3_s4_z}. In the lefthand column we plot the ratio $S^{ReBEL}_3/S^{\lcdm}_3$ of the skewness  
in the three different ReBEL models to the one for the canonical $\lcdm$ model in a sequence of five 
panels, for the five subsequent redshift snapshots which we analyzed, from $z=5.$ (bottom) to $z=0$ (top). 
The righthand column is organized in an equivalent manner for the kurtosis ratio $S^{ReBEL}_4/S^{\lcdm}_4$. 
In the panels we follow the same nomenclature and line scheme as in the previous section(s): the canonical 180LCDM 
$\lcdm$ model is indicated by the black solid line, the 180B-05RS1 ReBEL model with $\beta = -0.5$ and $r_s=1\hmpc$ 
by the blue dotted line, the 180B02RS1 ReBEL model with $\beta = 0.2$ and $r_s=1\hmpc$ by the green dot-dashed line 
and the 180B1RS1 ReBEL model with $\beta = 1.0$, $r_s=1\hmpc$ by the red dashed line. Also, we indicate the 
Nyquist scale $R\sim 1.4\hmpc$ again by means of the vertical dashed line. 

\begin{figure*}[t]
\mbox{\hskip -0.1truecm\includegraphics[width=0.35\textwidth,angle=-90]{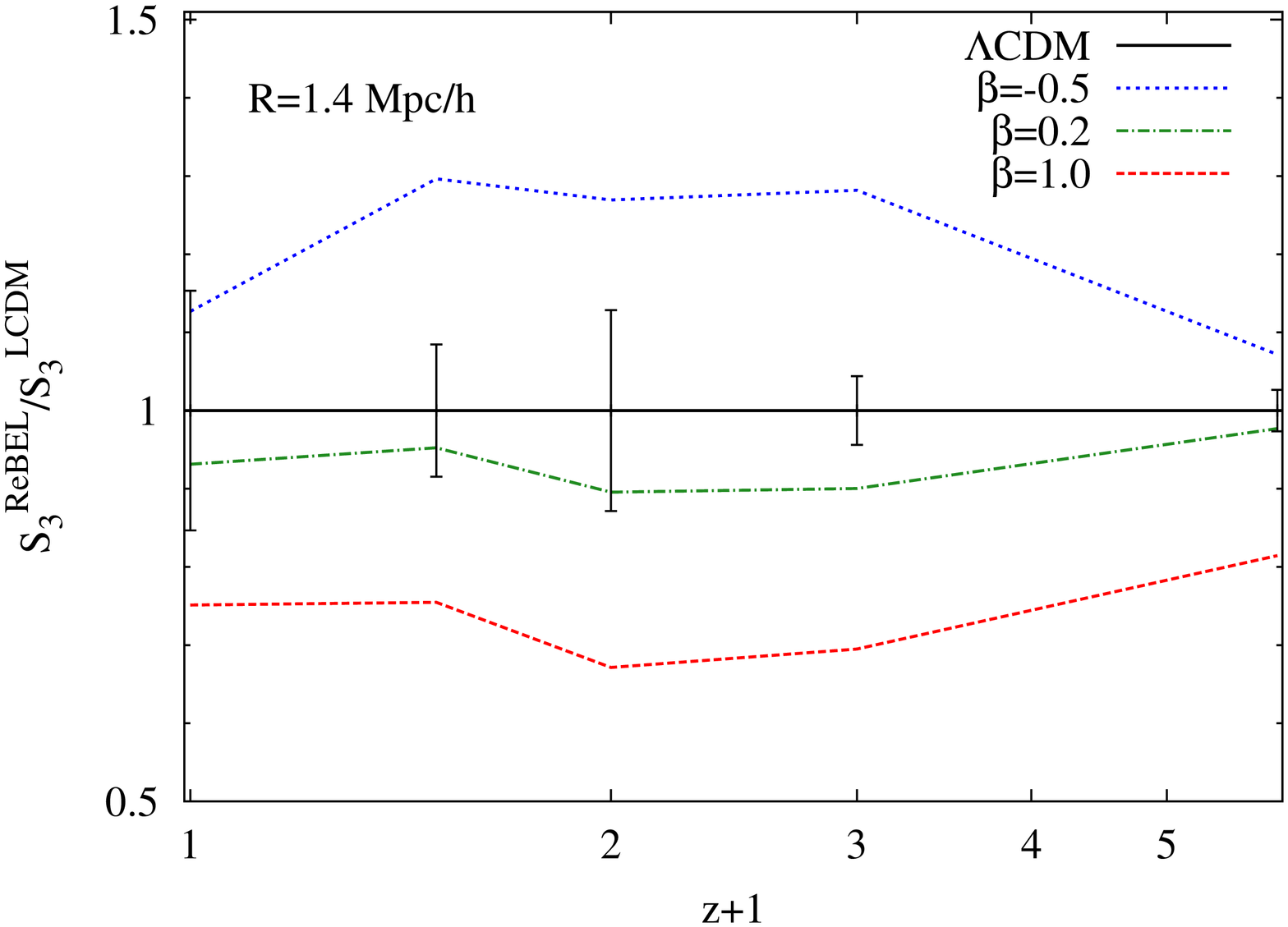}}
\includegraphics[width=0.35\textwidth,angle=-90]{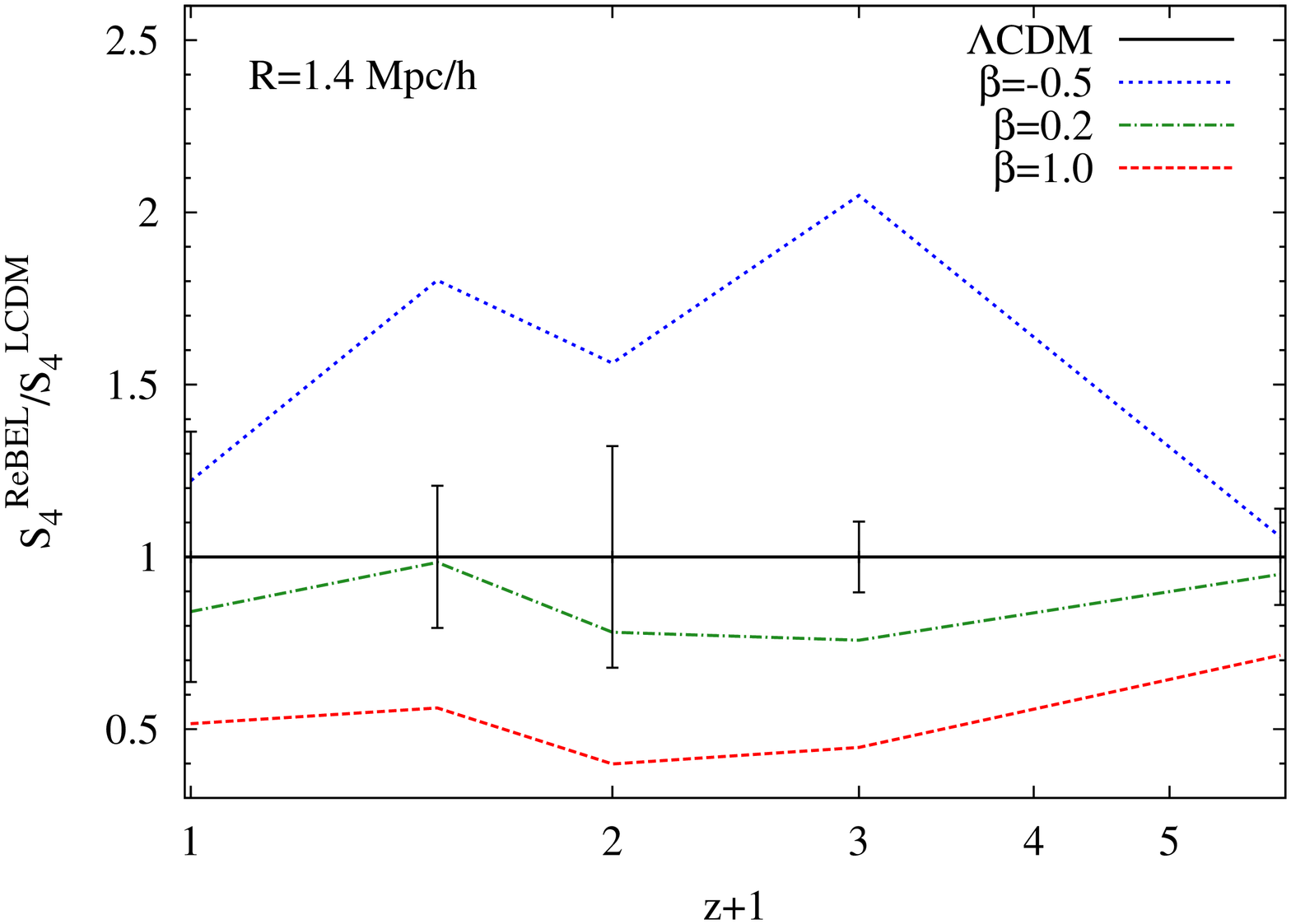}
\caption{Skewness ratio $S^{ReBEL}_3/S^{\lcdm}_3$ (lefthand panel) and kurtosis ratio 
$S^{ReBEL}_4/S^{\lcdm}_4$ (righthand panel) as a function of redshift $z$. Plotted are the ratios for 
three different ReBEL models: 180LCDM (solid line), 180B-05RS1(dashed line), 180B02RS1 (dotted line) and 
180B1RS1(dotted-dashed line).The error-bars correspond to the $1\sigma$ scatter of the 
180LCDM ensemble. 
\label{fig:s3_s4_allz_z}}
\end{figure*}
\begin{figure}[b]
\mbox{\hskip -0.1truecm\includegraphics[width=0.35\textwidth,angle=-90]{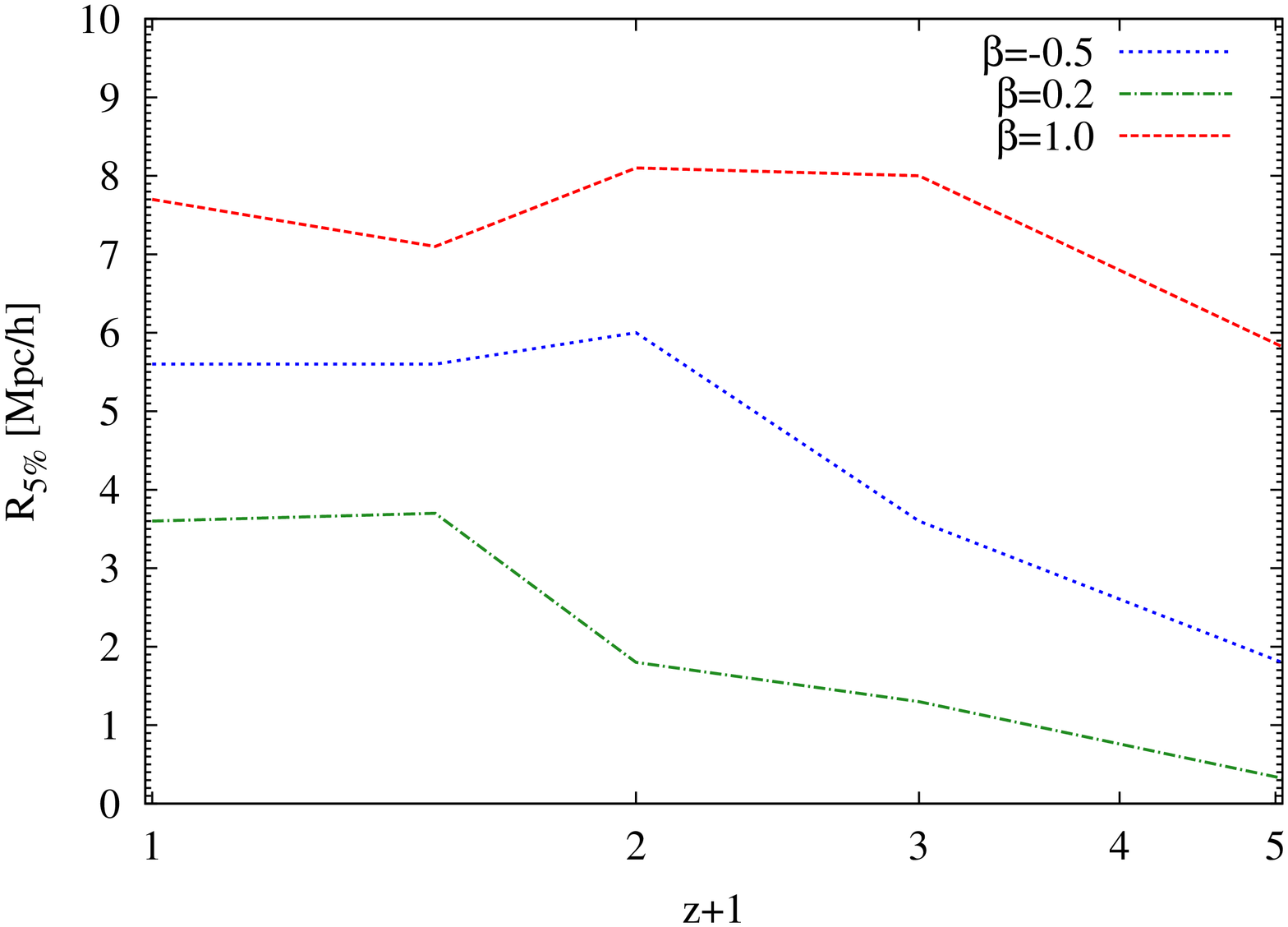}}
\caption{Evolution of skewness deviation scale $R_{5\%}$ as a function of redshift. At the scale $R_{5\%}$ 
the skewness for the ReBEL model deviates by $5\%$ from the value for the canonical $\lcdm$ model. Plotted 
are the deviation scales for three different ReBEL models: 180LCDM (solid line), 180B-05RS1(dashed line), 
180B02RS1 (dotted line) and 180B1RS1(dotted-dashed line).}
\label{fig:s3_rdep_z}
\end{figure}

\subsubsection{Deviation Scale}
Earlier, we had noted that at $z=0$ the ratio of the hierarchical amplitudes $S_n(R)/S_n^{\lcdm}$ in the 
various ReBEL models to that in the $\lcdm$ model is close to unity on large scales, scales considerably in 
excess of the ReBEL scale parameter $r_s$ and in the order of the scale of transition between linear and 
nonlinear evolution. When assessing this ratio for skewness and kurtosis at other redshifts, we notice the 
same trend. 

Interestingly, there is a slight but seemingly systematic shift in the scale at which the skewness 
and kurtosis ratios start to deviate significantly from unity. We observe that this scale gradually 
shifts towards larger scale as the evolution proceeds. When looking at the scale $R_{5\%}$ at which the skewness 
of the ReBEL models differs more than $\sim 5\%$ from the $\lcdm$ skewness, in the case 
of the $\beta=1.0$ ReBEL model we find that  at $z=5\hmpc$ it is only $R\sim 6 \hmpc$ while at $z=2$ it has 
increased to $R\sim 10\hmpc$ (see fig.~\ref{fig:s3_rdep_z}). The observed trend is directly linked to the 
scales on which the density field reaches non-linearity: the hierarchical amplitudes can only start 
to deviate from the canonical $\lcdm$ values through the related strong mode couplings. 
The other ReBEL models display similar evolutionary trends, although the details may differ somewhat.

At more recent redshifts, in all ReBEL models the growth of the deviations slows down, and at $z=0$ the scale is 
still $R_{5\%}\sim 10\hmpc$. Despite the growing amplitude of fluctuations at small nonlinear scales and the 
corresponding deviations of the ReBEL moments at these scales, the dynamical screening mechanism does 
not lead to the spread of these deviations to scales larger than $\sim 10\hmpc$. We may expect this, since the 
dynamical impact of the additional ReBEL scalar force will be rendered insignificant for Fourier modes smaller 
than the comoving Fourier mode $k_s=2\pi/r_s$\cite{NGP}. The required strong mode coupling 
will therefore not materialize. This observation is in agreement with the behaviour of the power 
spectrum $P(k)$ of the density perturbations, as noted in \cite{LRSI1,Rebel}. Figure~\ref{fig:s3_rdep_z} 
illustrates the convergence of the deviation scale $R_{5\%}$ in the case of all three ReBEL models. 

\subsubsection{Redshift Dependence}
Another interesting question with respect to the deviations of the ReBEL model skewness and kurtosis 
from the $\lcdm$ models concerns the issue at which redshift these are expected to be optimal. 
To address this issue, we assess the ReBEL $S_3$ and $S_4$ deviations on a scale of $\sim 1.4\hmpc$ scale. 
At this scale the find the highest deviations within the range set by the Nyquist scale. 

In table~\ref{tab:s3s4_z} we list the values of the skewness and kurtosis deviations $\Delta S_3$ and 
$\Delta S_4$, determined for the $180\hmpc$ ReBEL simulation ensembles 180B-05RS1, 180B02RS1 and 
180B1RS1. The magnitude of the hierarchy amplitude deviations $\Delta S_n$ is defined as:
\beq
\label{eqn:dev_def}
\Delta S_n = \left({S^{ReBEL}_n\over S^{\lcdm}_n} - 1\right)\,.
\eeq

Interestingly, the most pronounced discrepancies between the ReBEL and the standard model DM skewness and kurtosis 
are not found at the present epoch $z=0$. Instead, we find the maximal deviations in the range $0.5<z<2.0$. This 
is directly confirmed by the visual inspection of the two panels in figure~\ref{fig:s3_s4_allz_z}, where we 
plotted $1+\Delta S_3$ and $1+\Delta S_4$, ie. the ratios $S_3^{ReBEL}/S_3^{\lcdm}$ and $S_4^{ReBEL}/S_4^{\lcdm}$, 
versus redshift $z$. Amongst the rather limited redshift archive at our disposal, the maximum appears to be found 
at $z=1$. For a more precise determination of this epoch, we would need a considerably more densely binned 
redshift archive. Nonetheless, taking into account the errors in the amplitude estimates, we may confidently 
locate the maximum somewhere in the range quoted at the beginning of this paragraph.  

In our numerical experiments, at $z=1$ the $\Delta S_3$ reaches $10.3\%$ for the 180B02RS1 ensemble and $32.8\%$ 
for the 180B1RS1 ensemble. The $S_4$ deviations are considerably larger, and attain values of $21.9\%$ and $60.3\%$ 
for the same ensembles. We should emphasize that even while the deviations appear to reach their maximum at 
around $z=1$, they are still substantial at the current epoch, attaining values of $24.7\%$ and $48.4\%$ for the 
skewness and the kurtosis in the 180B1RS1 ReBEL ensemble. 

We therefore reach the important conclusion that the sharpest ``fingerprint'' of the scalar dark matter 
interactions, in terms of skewness and kurtosis, should be found at moderately intermediate redshifts. 

The answer to the question if these signatures can be detected in the observational reality depends on a range of 
issues. One of the most important ones is that of the bias between the dark matter distribution and the 
baryon density and galaxy distribution. As yet, there is not a definitive insight into how much this will be 
influenced by ReBEL dark matter forces. Keselman, Nusser and Peebles\cite{Rebel} recently studied the growth of 
cosmic structure in a simulation containing dark matter and baryons. While they confirm the expectation that the 
effect of ReBEL forces on the small scale baryon distribution is much smaller than that on dark matter, they 
also find that cannot be ignored. Preliminary results of our high-resolution joint simulation of baryons 
and DM within the ReBEL model (work in preparation) shows that the impact of the scalar interactions is not 
only imprinted on the moments of the dark matter density field, but also on the baryon density field. 
The implication may be that it will indeed be feasible to put observational constraints on the ReBEL 
cosmology parameter space with the help of galaxy catalogues.

\section{Discussion}
\label{sec-conclusions}
In this paper we have addressed the question in how far the differences between cosmological models 
involving a scalar long-range dark matter interaction would distinguish themselves from the canonical 
$\lcdm$ models in terms of their statistical properties. In answering this question, we focussed 
on the hierarchy of correlation functions and moments of the cosmological density fields. 

On the basis of a large ensemble of cosmological N-body simulations within the context of the 
standard $\lcdm$ cosmology and a range of different ReBEL long-range interaction models, we have 
attempted to identify the statistical differences between the models and the parameters and 
circumstances which will optimize our ability to discriminate between the different cosmologies. 
The simulations in this study are restricted to the pure dark matter distribution. 

To measure the moment and correlation functions, we base ourselves on the cumulants of the 
counts in cells of the particle distributions produced by the various N-body simulations. In the 
first stage of our study, we have thoroughly checked the accuracy and reliability of our implementation 
of the Counts in Cells method. We also assessed the influence of practical limitations, such as that 
of the finite length of a simulation box, on the measurements of the moments. This is particularly 
crucial given the circumstance that the gravitational force in the ReBEL models includes an 
intrinsic scale length. The conclusion from our experiments is that our implementation of the 
CIC method succesfully recovers the known results from perturbation theory in the context of the 
canonical $\lcdm$ model. 

Subsequently, we have applied our toolbox to the measurement of higher order N-point correlation 
functions, from $N=2$ to $N=9$, and the related hierarchical amplitudes, from the skewness $S_3$ and 
kurtosis $S_4$ to $S_8$. Amongst the most outstanding conclusions are:
\begin{itemize}
\item[$\bullet$] At scales comparable to the screening length parameter $r_s$ of the ReBEL model, the 
N-point functions $\bar \xi_N$ and the hierarchical amplitudes $S_n$ show deviations from the values 
expected in the standard $\lcdm$ cosmology.
\item[$\bullet$] In general, the amplitudes $S_N$ become smaller as the DM force strength parameter 
$\beta$ is larger. In the hypothetical situation of a negative $\beta$, the $S_n$ are larger than in the 
case of the $\lcdm$ cosmology. Usually, the magnitude of the $\beta=0.2$ model values $S_n$ are still in the 
order of the $\lcdm$ values, which technically correspond to the $\beta=0$ values. 
\item[$\bullet$] In a detailed comparison between the skewness and kurtosis of dark matter density fields, 
we find that the relative deviation of the kurtosis in the ReBEL models from that in the $\lcdm$ model 
is considerably larger than that for the skewness. In general, this is true for the whole range of 
hierarchical amplitudes: the deviation of $S_n$ in the ReBEL models is larger when it concerns a higher 
order $n$. 
\item[$\bullet$] The deviations of the hierarchical amplitudes $S_n(R)$ in the ReBEL models from that in  
the $\lcdm$ model are larger at smaller scales $R$. At scales where the evolution of the density field 
is still in the linear or quasi-linear regime, the deviations are negligible. Only at highly nonlinear 
scales we notice substantial and measurable differences. 
\item[$\bullet$] The scale at which we find substantial differences between $S_n(R)$ in the ReBEL 
models and in the canonical $\lcdm$ model gradually grows in time, a direct manifestation of the 
increasing scale of non-linearity as a result of cosmological structure growth. However, this increase comes 
to a halt when nonlinear structure growth has proceeded towards scales where the intrinsic screening length 
of the ReBEL forces calls a halt to its impact on the dark matter distribution. 
\item[$\bullet$] The deviations of the skewness $S_3(z)$ and kurtosis $S_4(z)$ of the ReBEL model from those 
in the $\lcdm$ cosmology reach their maximum in the moderate redshift range $0.5<z<2$. In other words, the 
imprint of ReBEL forces in the N-point correlation functions should be expected to be more prominent at 
medium redshift than at the current epoch. 
\end{itemize}
By confirming that there are noticeable differences in the higher-order clustering patterns between the standard 
$\lcdm$ cosmology and that in the ReBEL long-range dark matter interaction cosmologies we have identified a 
viable path towards constraining or falsifying these models on the basis of the observed galaxy distribution 
at moderate redshifts. Nonetheless, to be able to substantiate these claims we need to extend this analysis 
to more elaborate models. First, we need to assess whether the same significant conclusions may be drawn 
when the density field is sampled on the basis of dark matter halos and galaxies. This is a 
particularly important issue as the small measured differences between the LCDM and 
the ReBEL models might be washed out in the observationally relevant situation where the estimates are inferred from 
the dark matter halo distribution. Also, we need to investigate the extent to which these findings are influenced 
by working in redshift space instead of in regular (comoving) physical space. We foresee a substantial impact of the 
short-range ReBEL forces. 

In our upcoming study, we will address these questions on the basis of mock galaxy survey models, which 
will allow a direct comparison with circumstances prevailing in the observational reality.  

\acknowledgments
The authors would like to thank the anonymous referee for a careful appraisal which helped to 
significantly improve the content of this article. This research was partially supported by the Polish 
Ministry of Science Grant no. NN203 394234 and NN203 386037. The authors would like to thank Erwin Platen, 
Pawe\l{} Cieciel\c{a}g, Radek Wojtak and Micha\l{} Chodorowski for valuable discussions and comments. 
WAH acknowledges ASTROSIM exchange grant 2979 for enabling the extended workvisit to the Kapteyn Institute 
at the finishing stage of this paper, and WAH and RJ are grateful to NOVA for the NOVA visitor grant that 
started the collaboration at an earlier stage. WAH would like also to thanks the Kapteyn Institute for outstanding hospitality he received during his stay there. Simulations presented in this work were performed on the 'psk' cluster 
at Nicolaus Copernicus Astronomical Center and on the 'halo' and the 'halo2' clusters at Warsaw University Interdisciplinary Center 
for Mathematical and computational Modeling.

 \bibliography{scalar-pap-II-04}

\begin{thebibliography}{55}
\expandafter\ifx\csname natexlab\endcsname\relax\def\natexlab#1{#1}\fi
\expandafter\ifx\csname bibnamefont\endcsname\relax
  \def\bibnamefont#1{#1}\fi
\expandafter\ifx\csname bibfnamefont\endcsname\relax
  \def\bibfnamefont#1{#1}\fi
\expandafter\ifx\csname citenamefont\endcsname\relax
  \def\citenamefont#1{#1}\fi
\expandafter\ifx\csname url\endcsname\relax
  \def\url#1{\texttt{#1}}\fi
\expandafter\ifx\csname urlprefix\endcsname\relax\def\urlprefix{URL }\fi
\providecommand{\bibinfo}[2]{#2}
\providecommand{\eprint}[2][]{\url{#2}}

\bibitem[{\citenamefont{{Peebles}}(1980)}]{1980Peebles}
\bibinfo{author}{\bibfnamefont{P.~J.~E.} \bibnamefont{{Peebles}}},
  \emph{\bibinfo{title}{{The large-scale structure of the universe}}}
  (\bibinfo{publisher}{Research supported by the National Science
  Foundation.~Princeton, N.J., Princeton University Press, 1980.~435 p.},
  \bibinfo{year}{1980}).

\bibitem[{\citenamefont{{Juszkiewicz} et~al.}(1993)\citenamefont{{Juszkiewicz},
  {Bouchet}, and {Colombi}}}]{Juszkiewicz1993}
\bibinfo{author}{\bibfnamefont{R.}~\bibnamefont{{Juszkiewicz}}},
  \bibinfo{author}{\bibfnamefont{F.~R.} \bibnamefont{{Bouchet}}},
  \bibnamefont{and}
  \bibinfo{author}{\bibfnamefont{S.}~\bibnamefont{{Colombi}}},
  \bibinfo{journal}{\apj} \textbf{\bibinfo{volume}{412}}, \bibinfo{pages}{L9}
  (\bibinfo{year}{1993}), \eprint{arXiv:astro-ph/9306003}.

\bibitem[{\citenamefont{{Bernardeau}}(1992)}]{ber1992}
\bibinfo{author}{\bibfnamefont{F.}~\bibnamefont{{Bernardeau}}},
  \bibinfo{journal}{\apj} \textbf{\bibinfo{volume}{392}}, \bibinfo{pages}{1}
  (\bibinfo{year}{1992}).

\bibitem[{\citenamefont{{Szapudi} et~al.}(1999)\citenamefont{{Szapudi},
  {Quinn}, {Stadel}, and {Lake}}}]{npoint_omega_cdm}
\bibinfo{author}{\bibfnamefont{I.}~\bibnamefont{{Szapudi}}},
  \bibinfo{author}{\bibfnamefont{T.}~\bibnamefont{{Quinn}}},
  \bibinfo{author}{\bibfnamefont{J.}~\bibnamefont{{Stadel}}}, \bibnamefont{and}
  \bibinfo{author}{\bibfnamefont{G.}~\bibnamefont{{Lake}}},
  \bibinfo{journal}{\apj} \textbf{\bibinfo{volume}{517}}, \bibinfo{pages}{54}
  (\bibinfo{year}{1999}), \eprint{arXiv:astro-ph/9810190}.

\bibitem[{\citenamefont{{Durrer} et~al.}(2000)\citenamefont{{Durrer},
  {Juszkiewicz}, {Kunz}, and {Uzan}}}]{skew_nongaussian}
\bibinfo{author}{\bibfnamefont{R.}~\bibnamefont{{Durrer}}},
  \bibinfo{author}{\bibfnamefont{R.}~\bibnamefont{{Juszkiewicz}}},
  \bibinfo{author}{\bibfnamefont{M.}~\bibnamefont{{Kunz}}}, \bibnamefont{and}
  \bibinfo{author}{\bibfnamefont{J.}~\bibnamefont{{Uzan}}},
  \bibinfo{journal}{\prd} \textbf{\bibinfo{volume}{62}},
  \bibinfo{pages}{021301} (\bibinfo{year}{2000}),
  \eprint{arXiv:astro-ph/0005087}.

\bibitem[{\citenamefont{{Gaztanaga} and {Baugh}}(1995)}]{cic_nbody1}
\bibinfo{author}{\bibfnamefont{E.}~\bibnamefont{{Gaztanaga}}} \bibnamefont{and}
  \bibinfo{author}{\bibfnamefont{C.~M.} \bibnamefont{{Baugh}}},
  \bibinfo{journal}{\mnras} \textbf{\bibinfo{volume}{273}}, \bibinfo{pages}{L1}
  (\bibinfo{year}{1995}), \eprint{arXiv:astro-ph/9409062}.

\bibitem[{\citenamefont{{Bouchet} and {Hernquist}}(1992)}]{cic_nbody2}
\bibinfo{author}{\bibfnamefont{F.~R.} \bibnamefont{{Bouchet}}}
  \bibnamefont{and}
  \bibinfo{author}{\bibfnamefont{L.}~\bibnamefont{{Hernquist}}},
  \bibinfo{journal}{\apj} \textbf{\bibinfo{volume}{400}}, \bibinfo{pages}{25}
  (\bibinfo{year}{1992}).

\bibitem[{\citenamefont{{Farrar} and
  {Peebles}}(2004{\natexlab{a}})}]{FarrarPeebles}
\bibinfo{author}{\bibfnamefont{G.~R.} \bibnamefont{{Farrar}}} \bibnamefont{and}
  \bibinfo{author}{\bibfnamefont{P.~J.~E.} \bibnamefont{{Peebles}}},
  \bibinfo{journal}{Astrophys.~J.} \textbf{\bibinfo{volume}{604}},
  \bibinfo{pages}{1} (\bibinfo{year}{2004}{\natexlab{a}}),
  \eprint{arXiv:astro-ph/0307316}.

\bibitem[{\citenamefont{{Farrar} and
  {Peebles}}(2004{\natexlab{b}})}]{PeeblesIDMDE}
\bibinfo{author}{\bibfnamefont{G.~R.} \bibnamefont{{Farrar}}} \bibnamefont{and}
  \bibinfo{author}{\bibfnamefont{P.~J.~E.} \bibnamefont{{Peebles}}},
  \bibinfo{journal}{Astrophys.~J.} \textbf{\bibinfo{volume}{604}},
  \bibinfo{pages}{1} (\bibinfo{year}{2004}{\natexlab{b}}),
  \eprint{arXiv:astro-ph/0307316}.

\bibitem[{\citenamefont{{Gubser} and {Peebles}}(2004{\natexlab{a}})}]{GP1}
\bibinfo{author}{\bibfnamefont{S.~S.} \bibnamefont{{Gubser}}} \bibnamefont{and}
  \bibinfo{author}{\bibfnamefont{P.~J.~E.} \bibnamefont{{Peebles}}},
  \bibinfo{journal}{Phys.~Rev.~D} \textbf{\bibinfo{volume}{70}},
  \bibinfo{pages}{123510} (\bibinfo{year}{2004}{\natexlab{a}}),
  \eprint{arXiv:hep-th/0402225}.

\bibitem[{\citenamefont{{Gubser} and {Peebles}}(2004{\natexlab{b}})}]{GP2}
\bibinfo{author}{\bibfnamefont{S.~S.} \bibnamefont{{Gubser}}} \bibnamefont{and}
  \bibinfo{author}{\bibfnamefont{P.~J.~E.} \bibnamefont{{Peebles}}},
  \bibinfo{journal}{Phys.~Rev.~D} \textbf{\bibinfo{volume}{70}},
  \bibinfo{pages}{123511} (\bibinfo{year}{2004}{\natexlab{b}}),
  \eprint{arXiv:hep-th/0407097}.

\bibitem[{\citenamefont{{Farrar} and {Rosen}}(2007)}]{Farrar2007}
\bibinfo{author}{\bibfnamefont{G.~R.} \bibnamefont{{Farrar}}} \bibnamefont{and}
  \bibinfo{author}{\bibfnamefont{R.~A.} \bibnamefont{{Rosen}}},
  \bibinfo{journal}{Phys.~Rev.~Lett.} \textbf{\bibinfo{volume}{98}},
  \bibinfo{pages}{171302} (\bibinfo{year}{2007}),
  \eprint{arXiv:astro-ph/0610298}.

\bibitem[{\citenamefont{{Brookfield} et~al.}(2008)\citenamefont{{Brookfield},
  {van de Bruck}, and {Hall}}}]{Brookfield}
\bibinfo{author}{\bibfnamefont{A.~W.} \bibnamefont{{Brookfield}}},
  \bibinfo{author}{\bibfnamefont{C.}~\bibnamefont{{van de Bruck}}},
  \bibnamefont{and} \bibinfo{author}{\bibfnamefont{L.~M.~H.}
  \bibnamefont{{Hall}}}, \bibinfo{journal}{Phys.~Rev.~D}
  \textbf{\bibinfo{volume}{77}}, \bibinfo{pages}{043006}
  (\bibinfo{year}{2008}), \eprint{arXiv:0709.2297}.

\bibitem[{\citenamefont{{Peebles}}(2001)}]{PeeblesVoid}
\bibinfo{author}{\bibfnamefont{P.~J.~E.} \bibnamefont{{Peebles}}},
  \bibinfo{journal}{Astrophys.~J.} \textbf{\bibinfo{volume}{557}},
  \bibinfo{pages}{495} (\bibinfo{year}{2001}), \eprint{arXiv:astro-ph/0101127}.

\bibitem[{\citenamefont{{Peebles}}(2010)}]{peeb2}
\bibinfo{author}{\bibfnamefont{P.~J.~E.} \bibnamefont{{Peebles}}}, in
  \emph{\bibinfo{booktitle}{American Institute of Physics Conference Series}},
  edited by \bibinfo{editor}{\bibnamefont{{J.-M.~Alimi \& A.~Fu{\"o}zfa}}}
  (\bibinfo{year}{2010}), vol. \bibinfo{volume}{1241} of
  \emph{\bibinfo{series}{American Institute of Physics Conference Series}}, pp.
  \bibinfo{pages}{175--182}, \eprint{0910.5142}.

\bibitem[{\citenamefont{{Peebles} and {Nusser}}(2010)}]{peeb1}
\bibinfo{author}{\bibfnamefont{P.~J.~E.} \bibnamefont{{Peebles}}}
  \bibnamefont{and} \bibinfo{author}{\bibfnamefont{A.}~\bibnamefont{{Nusser}}},
  \bibinfo{journal}{\nat} \textbf{\bibinfo{volume}{465}}, \bibinfo{pages}{565}
  (\bibinfo{year}{2010}), \eprint{1001.1484}.

\bibitem[{\citenamefont{{Nusser} et~al.}(2005)\citenamefont{{Nusser}, {Gubser},
  and {Peebles}}}]{NGP}
\bibinfo{author}{\bibfnamefont{A.}~\bibnamefont{{Nusser}}},
  \bibinfo{author}{\bibfnamefont{S.~S.} \bibnamefont{{Gubser}}},
  \bibnamefont{and} \bibinfo{author}{\bibfnamefont{P.~J.}
  \bibnamefont{{Peebles}}}, \bibinfo{journal}{Phys.~Rev.~D}
  \textbf{\bibinfo{volume}{71}}, \bibinfo{pages}{083505}
  (\bibinfo{year}{2005}), \eprint{arXiv:astro-ph/0412586}.

\bibitem[{\citenamefont{{Hellwing} and {Juszkiewicz}}(2009)}]{LRSI1}
\bibinfo{author}{\bibfnamefont{W.~A.} \bibnamefont{{Hellwing}}}
  \bibnamefont{and}
  \bibinfo{author}{\bibfnamefont{R.}~\bibnamefont{{Juszkiewicz}}},
  \bibinfo{journal}{Phys.~Rev.~D} \textbf{\bibinfo{volume}{80}},
  \bibinfo{pages}{083522} (\bibinfo{year}{2009}), \eprint{0809.1976}.

\bibitem[{\citenamefont{{Hellwing}}(2010)}]{LRSI2}
\bibinfo{author}{\bibfnamefont{W.~A.} \bibnamefont{{Hellwing}}},
  \bibinfo{journal}{Annalen der Physik} \textbf{\bibinfo{volume}{19}},
  \bibinfo{pages}{351} (\bibinfo{year}{2010}), \eprint{0911.0573}.

\bibitem[{\citenamefont{{Keselman} et~al.}(2010)\citenamefont{{Keselman},
  {Nusser}, and {Peebles}}}]{Rebel}
\bibinfo{author}{\bibfnamefont{J.~A.} \bibnamefont{{Keselman}}},
  \bibinfo{author}{\bibfnamefont{A.}~\bibnamefont{{Nusser}}}, \bibnamefont{and}
  \bibinfo{author}{\bibfnamefont{P.~J.~E.} \bibnamefont{{Peebles}}},
  \bibinfo{journal}{\prd} \textbf{\bibinfo{volume}{81}},
  \bibinfo{pages}{063521} (\bibinfo{year}{2010}), \eprint{0912.4177}.

\bibitem[{\citenamefont{{Hellwing} et~al.}(2010)\citenamefont{{Hellwing},
  {Knollmann}, and {Knebe}}}]{Rebel2}
\bibinfo{author}{\bibfnamefont{W.~A.} \bibnamefont{{Hellwing}}},
  \bibinfo{author}{\bibfnamefont{S.~R.} \bibnamefont{{Knollmann}}},
  \bibnamefont{and} \bibinfo{author}{\bibfnamefont{A.}~\bibnamefont{{Knebe}}},
  \bibinfo{journal}{\mnras} \textbf{\bibinfo{volume}{408}},
  \bibinfo{pages}{L104} (\bibinfo{year}{2010}), \eprint{1004.2929}.

\bibitem[{\citenamefont{{Keselman} et~al.}(2009)\citenamefont{{Keselman},
  {Nusser}, and {Peebles}}}]{reb_sat1}
\bibinfo{author}{\bibfnamefont{J.~A.} \bibnamefont{{Keselman}}},
  \bibinfo{author}{\bibfnamefont{A.}~\bibnamefont{{Nusser}}}, \bibnamefont{and}
  \bibinfo{author}{\bibfnamefont{P.~J.~E.} \bibnamefont{{Peebles}}},
  \bibinfo{journal}{\prd} \textbf{\bibinfo{volume}{80}},
  \bibinfo{pages}{063517} (\bibinfo{year}{2009}), \eprint{0902.3452}.

\bibitem[{\citenamefont{{Kesden}}(2009)}]{reb_sat2}
\bibinfo{author}{\bibfnamefont{M.}~\bibnamefont{{Kesden}}},
  \bibinfo{journal}{\prd} \textbf{\bibinfo{volume}{80}},
  \bibinfo{pages}{083530} (\bibinfo{year}{2009}), \eprint{0903.4458}.

\bibitem[{\citenamefont{{Baldi} et~al.}(2010)\citenamefont{{Baldi},
  {Pettorino}, {Robbers}, and {Springel}}}]{Coupled_DE1}
\bibinfo{author}{\bibfnamefont{M.}~\bibnamefont{{Baldi}}},
  \bibinfo{author}{\bibfnamefont{V.}~\bibnamefont{{Pettorino}}},
  \bibinfo{author}{\bibfnamefont{G.}~\bibnamefont{{Robbers}}},
  \bibnamefont{and}
  \bibinfo{author}{\bibfnamefont{V.}~\bibnamefont{{Springel}}},
  \bibinfo{journal}{\mnras} \textbf{\bibinfo{volume}{403}},
  \bibinfo{pages}{1684} (\bibinfo{year}{2010}), \eprint{0812.3901}.

\bibitem[{\citenamefont{{Baldi}}(2009)}]{Coupled_DE2}
\bibinfo{author}{\bibfnamefont{M.}~\bibnamefont{{Baldi}}},
  \bibinfo{journal}{Nuclear Physics B Proceedings Supplements}
  \textbf{\bibinfo{volume}{194}}, \bibinfo{pages}{178} (\bibinfo{year}{2009}),
  \eprint{0906.5353}.

\bibitem[{\citenamefont{{Li} and {Barrow}}(2010{\natexlab{a}})}]{Coupled_DE3}
\bibinfo{author}{\bibfnamefont{B.}~\bibnamefont{{Li}}} \bibnamefont{and}
  \bibinfo{author}{\bibfnamefont{J.~D.} \bibnamefont{{Barrow}}},
  \bibinfo{journal}{ArXiv e-prints}  (\bibinfo{year}{2010}{\natexlab{a}}),
  \eprint{1005.4231}.

\bibitem[{\citenamefont{{Li} and {Zhao}}(2010)}]{Coupled_DE4}
\bibinfo{author}{\bibfnamefont{B.}~\bibnamefont{{Li}}} \bibnamefont{and}
  \bibinfo{author}{\bibfnamefont{H.}~\bibnamefont{{Zhao}}},
  \bibinfo{journal}{\prd} \textbf{\bibinfo{volume}{81}},
  \bibinfo{pages}{104047} (\bibinfo{year}{2010}), \eprint{1001.3152}.

\bibitem[{\citenamefont{{Li}}(2010)}]{Coupled_DE5}
\bibinfo{author}{\bibfnamefont{B.}~\bibnamefont{{Li}}}, \bibinfo{journal}{ArXiv
  e-prints}  (\bibinfo{year}{2010}), \eprint{1009.1406}.

\bibitem[{\citenamefont{{Li} et~al.}(2010{\natexlab{a}})\citenamefont{{Li},
  {Mota}, and {Barrow}}}]{Coupled_DE6}
\bibinfo{author}{\bibfnamefont{B.}~\bibnamefont{{Li}}},
  \bibinfo{author}{\bibfnamefont{D.~F.} \bibnamefont{{Mota}}},
  \bibnamefont{and} \bibinfo{author}{\bibfnamefont{J.~D.}
  \bibnamefont{{Barrow}}}, \bibinfo{journal}{ArXiv e-prints}
  (\bibinfo{year}{2010}{\natexlab{a}}), \eprint{1009.1400}.

\bibitem[{\citenamefont{{Li} et~al.}(2010{\natexlab{b}})\citenamefont{{Li},
  {Mota}, and {Barrow}}}]{Coupled_DE7}
\bibinfo{author}{\bibfnamefont{B.}~\bibnamefont{{Li}}},
  \bibinfo{author}{\bibfnamefont{D.~F.} \bibnamefont{{Mota}}},
  \bibnamefont{and} \bibinfo{author}{\bibfnamefont{J.~D.}
  \bibnamefont{{Barrow}}}, \bibinfo{journal}{ArXiv e-prints}
  (\bibinfo{year}{2010}{\natexlab{b}}), \eprint{1009.1396}.

\bibitem[{\citenamefont{{Li} and {Barrow}}(2010{\natexlab{b}})}]{COupled_DE8}
\bibinfo{author}{\bibfnamefont{B.}~\bibnamefont{{Li}}} \bibnamefont{and}
  \bibinfo{author}{\bibfnamefont{J.~D.} \bibnamefont{{Barrow}}},
  \bibinfo{journal}{ArXiv e-prints}  (\bibinfo{year}{2010}{\natexlab{b}}),
  \eprint{1010.3748}.

\bibitem[{\citenamefont{{Lee}}(2010)}]{lee}
\bibinfo{author}{\bibfnamefont{J.}~\bibnamefont{{Lee}}},
  \bibinfo{journal}{ArXiv e-prints}  (\bibinfo{year}{2010}),
  \eprint{1008.4620}.

\bibitem[{\citenamefont{{Klypin} and {Holtzman}}(1997)}]{PMcode}
\bibinfo{author}{\bibfnamefont{A.}~\bibnamefont{{Klypin}}} \bibnamefont{and}
  \bibinfo{author}{\bibfnamefont{J.}~\bibnamefont{{Holtzman}}},
  \bibinfo{journal}{ArXiv e-prints}  (\bibinfo{year}{1997}),
  \eprint{astro-ph/9712217}.

\bibitem[{\citenamefont{{Seljak} and {Zaldarriaga}}(1996)}]{cmbfast}
\bibinfo{author}{\bibfnamefont{U.}~\bibnamefont{{Seljak}}} \bibnamefont{and}
  \bibinfo{author}{\bibfnamefont{M.}~\bibnamefont{{Zaldarriaga}}},
  \bibinfo{journal}{Astrophys.~J.} \textbf{\bibinfo{volume}{469}},
  \bibinfo{pages}{437} (\bibinfo{year}{1996}), \eprint{arXiv:astro-ph/9603033}.

\bibitem[{\citenamefont{{Nesseris} and {Perivolaropoulos}}(2004)}]{SCDM1}
\bibinfo{author}{\bibfnamefont{S.}~\bibnamefont{{Nesseris}}} \bibnamefont{and}
  \bibinfo{author}{\bibfnamefont{L.}~\bibnamefont{{Perivolaropoulos}}},
  \bibinfo{journal}{\prd} \textbf{\bibinfo{volume}{70}},
  \bibinfo{pages}{043531} (\bibinfo{year}{2004}),
  \eprint{arXiv:astro-ph/0401556}.

\bibitem[{\citenamefont{{Governato} et~al.}(1999)\citenamefont{{Governato},
  {Babul}, {Quinn}, {Tozzi}, {Baugh}, {Katz}, and {Lake}}}]{SCDM2}
\bibinfo{author}{\bibfnamefont{F.}~\bibnamefont{{Governato}}},
  \bibinfo{author}{\bibfnamefont{A.}~\bibnamefont{{Babul}}},
  \bibinfo{author}{\bibfnamefont{T.}~\bibnamefont{{Quinn}}},
  \bibinfo{author}{\bibfnamefont{P.}~\bibnamefont{{Tozzi}}},
  \bibinfo{author}{\bibfnamefont{C.~M.} \bibnamefont{{Baugh}}},
  \bibinfo{author}{\bibfnamefont{N.}~\bibnamefont{{Katz}}}, \bibnamefont{and}
  \bibinfo{author}{\bibfnamefont{G.}~\bibnamefont{{Lake}}},
  \bibinfo{journal}{\mnras} \textbf{\bibinfo{volume}{307}},
  \bibinfo{pages}{949} (\bibinfo{year}{1999}), \eprint{arXiv:astro-ph/9810189}.

\bibitem[{\citenamefont{{White} et~al.}(1993)\citenamefont{{White},
  {Efstathiou}, and {Frenk}}}]{SCDM3}
\bibinfo{author}{\bibfnamefont{S.~D.~M.} \bibnamefont{{White}}},
  \bibinfo{author}{\bibfnamefont{G.}~\bibnamefont{{Efstathiou}}},
  \bibnamefont{and} \bibinfo{author}{\bibfnamefont{C.~S.}
  \bibnamefont{{Frenk}}}, \bibinfo{journal}{\mnras}
  \textbf{\bibinfo{volume}{262}}, \bibinfo{pages}{1023} (\bibinfo{year}{1993}).

\bibitem[{\citenamefont{{Bertschinger} and {Juszkiewicz}}(1988)}]{SCDM_GA}
\bibinfo{author}{\bibfnamefont{E.}~\bibnamefont{{Bertschinger}}}
  \bibnamefont{and}
  \bibinfo{author}{\bibfnamefont{R.}~\bibnamefont{{Juszkiewicz}}},
  \bibinfo{journal}{\apjl} \textbf{\bibinfo{volume}{334}}, \bibinfo{pages}{L59}
  (\bibinfo{year}{1988}).

\bibitem[{\citenamefont{{Feldman} et~al.}(2003)\citenamefont{{Feldman},
  {Juszkiewicz}, {Ferreira}, {Davis}, {Gazta{\~n}aga}, {Fry}, {Jaffe},
  {Chambers}, {da Costa}, {Bernardi} et~al.}}]{FeldmanOmega}
\bibinfo{author}{\bibfnamefont{H.}~\bibnamefont{{Feldman}}},
  \bibinfo{author}{\bibfnamefont{R.}~\bibnamefont{{Juszkiewicz}}},
  \bibinfo{author}{\bibfnamefont{P.}~\bibnamefont{{Ferreira}}},
  \bibinfo{author}{\bibfnamefont{M.}~\bibnamefont{{Davis}}},
  \bibinfo{author}{\bibfnamefont{E.}~\bibnamefont{{Gazta{\~n}aga}}},
  \bibinfo{author}{\bibfnamefont{J.}~\bibnamefont{{Fry}}},
  \bibinfo{author}{\bibfnamefont{A.}~\bibnamefont{{Jaffe}}},
  \bibinfo{author}{\bibfnamefont{S.}~\bibnamefont{{Chambers}}},
  \bibinfo{author}{\bibfnamefont{L.}~\bibnamefont{{da Costa}}},
  \bibinfo{author}{\bibfnamefont{M.}~\bibnamefont{{Bernardi}}},
  \bibnamefont{et~al.}, \bibinfo{journal}{\apjl}
  \textbf{\bibinfo{volume}{596}}, \bibinfo{pages}{L131} (\bibinfo{year}{2003}),
  \eprint{arXiv:astro-ph/0305078}.

\bibitem[{\citenamefont{{Riess} et~al.}(1998)\citenamefont{{Riess},
  {Filippenko}, {Challis}, {Clocchiatti}, {Diercks}, {Garnavich}, {Gilliland},
  {Hogan}, {Jha}, {Kirshner} et~al.}}]{acceleration1}
\bibinfo{author}{\bibfnamefont{A.~G.} \bibnamefont{{Riess}}},
  \bibinfo{author}{\bibfnamefont{A.~V.} \bibnamefont{{Filippenko}}},
  \bibinfo{author}{\bibfnamefont{P.}~\bibnamefont{{Challis}}},
  \bibinfo{author}{\bibfnamefont{A.}~\bibnamefont{{Clocchiatti}}},
  \bibinfo{author}{\bibfnamefont{A.}~\bibnamefont{{Diercks}}},
  \bibinfo{author}{\bibfnamefont{P.~M.} \bibnamefont{{Garnavich}}},
  \bibinfo{author}{\bibfnamefont{R.~L.} \bibnamefont{{Gilliland}}},
  \bibinfo{author}{\bibfnamefont{C.~J.} \bibnamefont{{Hogan}}},
  \bibinfo{author}{\bibfnamefont{S.}~\bibnamefont{{Jha}}},
  \bibinfo{author}{\bibfnamefont{R.~P.} \bibnamefont{{Kirshner}}},
  \bibnamefont{et~al.}, \bibinfo{journal}{\aj} \textbf{\bibinfo{volume}{116}},
  \bibinfo{pages}{1009} (\bibinfo{year}{1998}),
  \eprint{arXiv:astro-ph/9805201}.

\bibitem[{\citenamefont{{Perlmutter} et~al.}(1999)\citenamefont{{Perlmutter},
  {Aldering}, {Goldhaber}, {Knop}, {Nugent}, {Castro}, {Deustua}, {Fabbro},
  {Goobar}, {Groom} et~al.}}]{acceleration2}
\bibinfo{author}{\bibfnamefont{S.}~\bibnamefont{{Perlmutter}}},
  \bibinfo{author}{\bibfnamefont{G.}~\bibnamefont{{Aldering}}},
  \bibinfo{author}{\bibfnamefont{G.}~\bibnamefont{{Goldhaber}}},
  \bibinfo{author}{\bibfnamefont{R.~A.} \bibnamefont{{Knop}}},
  \bibinfo{author}{\bibfnamefont{P.}~\bibnamefont{{Nugent}}},
  \bibinfo{author}{\bibfnamefont{P.~G.} \bibnamefont{{Castro}}},
  \bibinfo{author}{\bibfnamefont{S.}~\bibnamefont{{Deustua}}},
  \bibinfo{author}{\bibfnamefont{S.}~\bibnamefont{{Fabbro}}},
  \bibinfo{author}{\bibfnamefont{A.}~\bibnamefont{{Goobar}}},
  \bibinfo{author}{\bibfnamefont{D.~E.} \bibnamefont{{Groom}}},
  \bibnamefont{et~al.}, \bibinfo{journal}{\apj} \textbf{\bibinfo{volume}{517}},
  \bibinfo{pages}{565} (\bibinfo{year}{1999}), \eprint{arXiv:astro-ph/9812133}.

\bibitem[{\citenamefont{{Baugh} et~al.}(1995)\citenamefont{{Baugh},
  {Gaztanaga}, and {Efstathiou}}}]{cic_ana}
\bibinfo{author}{\bibfnamefont{C.~M.} \bibnamefont{{Baugh}}},
  \bibinfo{author}{\bibfnamefont{E.}~\bibnamefont{{Gaztanaga}}},
  \bibnamefont{and}
  \bibinfo{author}{\bibfnamefont{G.}~\bibnamefont{{Efstathiou}}},
  \bibinfo{journal}{\mnras} \textbf{\bibinfo{volume}{274}},
  \bibinfo{pages}{1049} (\bibinfo{year}{1995}),
  \eprint{arXiv:astro-ph/9408057}.

\bibitem[{\citenamefont{{White} and {Frenk}}(1991)}]{White_SCDM}
\bibinfo{author}{\bibfnamefont{S.~D.~M.} \bibnamefont{{White}}}
  \bibnamefont{and} \bibinfo{author}{\bibfnamefont{C.~S.}
  \bibnamefont{{Frenk}}}, \bibinfo{journal}{\apj}
  \textbf{\bibinfo{volume}{379}}, \bibinfo{pages}{52} (\bibinfo{year}{1991}).

\bibitem[{\citenamefont{{Gaztanaga} and {Frieman}}(1994)}]{SCDM4}
\bibinfo{author}{\bibfnamefont{E.}~\bibnamefont{{Gaztanaga}}} \bibnamefont{and}
  \bibinfo{author}{\bibfnamefont{J.~A.} \bibnamefont{{Frieman}}},
  \bibinfo{journal}{\apjl} \textbf{\bibinfo{volume}{437}}, \bibinfo{pages}{L13}
  (\bibinfo{year}{1994}), \eprint{arXiv:astro-ph/9407079}.

\bibitem[{\citenamefont{{Springel}}(2005)}]{Gadget2}
\bibinfo{author}{\bibfnamefont{V.}~\bibnamefont{{Springel}}},
  \bibinfo{journal}{Mon.~Not.~Roy.~Astron.~Soc.}
  \textbf{\bibinfo{volume}{364}}, \bibinfo{pages}{1105} (\bibinfo{year}{2005}),
  \eprint{arXiv:astro-ph/0505010}.

\bibitem[{\citenamefont{{Gaztanaga}}(1994)}]{GaztanagaAPM94}
\bibinfo{author}{\bibfnamefont{E.}~\bibnamefont{{Gaztanaga}}},
  \bibinfo{journal}{\mnras} \textbf{\bibinfo{volume}{268}},
  \bibinfo{pages}{913} (\bibinfo{year}{1994}), \eprint{arXiv:astro-ph/9309019}.

\bibitem[{\citenamefont{{Bernardeau} et~al.}(2002)\citenamefont{{Bernardeau},
  {Colombi}, {Gazta{\~n}aga}, and {Scoccimarro}}}]{BCGS_book}
\bibinfo{author}{\bibfnamefont{F.}~\bibnamefont{{Bernardeau}}},
  \bibinfo{author}{\bibfnamefont{S.}~\bibnamefont{{Colombi}}},
  \bibinfo{author}{\bibfnamefont{E.}~\bibnamefont{{Gazta{\~n}aga}}},
  \bibnamefont{and}
  \bibinfo{author}{\bibfnamefont{R.}~\bibnamefont{{Scoccimarro}}},
  \bibinfo{journal}{\physrep} \textbf{\bibinfo{volume}{367}},
  \bibinfo{pages}{1} (\bibinfo{year}{2002}), \eprint{arXiv:astro-ph/0112551}.

\bibitem[{\citenamefont{{Szapudi} and {Colombi}}(1996)}]{cosmic_error}
\bibinfo{author}{\bibfnamefont{I.}~\bibnamefont{{Szapudi}}} \bibnamefont{and}
  \bibinfo{author}{\bibfnamefont{S.}~\bibnamefont{{Colombi}}},
  \bibinfo{journal}{\apj} \textbf{\bibinfo{volume}{470}}, \bibinfo{pages}{131}
  (\bibinfo{year}{1996}), \eprint{arXiv:astro-ph/9510030}.

\bibitem[{\citenamefont{{Bernardeau}}(1994)}]{skew_cur_pt}
\bibinfo{author}{\bibfnamefont{F.}~\bibnamefont{{Bernardeau}}},
  \bibinfo{journal}{\apj} \textbf{\bibinfo{volume}{433}}, \bibinfo{pages}{1}
  (\bibinfo{year}{1994}), \eprint{arXiv:astro-ph/9312026}.

\bibitem[{\citenamefont{{Colombi} et~al.}(1994)\citenamefont{{Colombi},
  {Bouchet}, and {Schaeffer}}}]{Colombi1994}
\bibinfo{author}{\bibfnamefont{S.}~\bibnamefont{{Colombi}}},
  \bibinfo{author}{\bibfnamefont{F.~R.} \bibnamefont{{Bouchet}}},
  \bibnamefont{and}
  \bibinfo{author}{\bibfnamefont{R.}~\bibnamefont{{Schaeffer}}},
  \bibinfo{journal}{\aap} \textbf{\bibinfo{volume}{281}}, \bibinfo{pages}{301}
  (\bibinfo{year}{1994}).

\bibitem[{\citenamefont{{Zel'Dovich}}(1970)}]{ZA}
\bibinfo{author}{\bibfnamefont{Y.~B.} \bibnamefont{{Zel'Dovich}}},
  \bibinfo{journal}{\aap} \textbf{\bibinfo{volume}{5}}, \bibinfo{pages}{84}
  (\bibinfo{year}{1970}).

\bibitem[{\citenamefont{{Crocce} et~al.}(2006)\citenamefont{{Crocce},
  {Pueblas}, and {Scoccimarro}}}]{transients1}
\bibinfo{author}{\bibfnamefont{M.}~\bibnamefont{{Crocce}}},
  \bibinfo{author}{\bibfnamefont{S.}~\bibnamefont{{Pueblas}}},
  \bibnamefont{and}
  \bibinfo{author}{\bibfnamefont{R.}~\bibnamefont{{Scoccimarro}}},
  \bibinfo{journal}{\mnras} \textbf{\bibinfo{volume}{373}},
  \bibinfo{pages}{369} (\bibinfo{year}{2006}), \eprint{arXiv:astro-ph/0606505}.

\bibitem[{\citenamefont{{Tatekawa} and {Mizuno}}(2007)}]{transients2}
\bibinfo{author}{\bibfnamefont{T.}~\bibnamefont{{Tatekawa}}} \bibnamefont{and}
  \bibinfo{author}{\bibfnamefont{S.}~\bibnamefont{{Mizuno}}},
  \bibinfo{journal}{Journal of Cosmology and Astro-Particle Physics}
  \textbf{\bibinfo{volume}{12}}, \bibinfo{pages}{14} (\bibinfo{year}{2007}),
  \eprint{0706.1334}.

\bibitem[{\citenamefont{{Scoccimarro}}(1998)}]{scoccimarro98}
\bibinfo{author}{\bibfnamefont{R.}~\bibnamefont{{Scoccimarro}}},
  \bibinfo{journal}{\mnras} \textbf{\bibinfo{volume}{299}},
  \bibinfo{pages}{1097} (\bibinfo{year}{1998}),
  \eprint{arXiv:astro-ph/9711187}.

\bibitem[{\citenamefont{{Jenkins}}(2010)}]{jenkins10}
\bibinfo{author}{\bibfnamefont{A.}~\bibnamefont{{Jenkins}}},
  \bibinfo{journal}{\mnras} \textbf{\bibinfo{volume}{403}},
  \bibinfo{pages}{1859} (\bibinfo{year}{2010}), \eprint{0910.0258}.

\end{thebibliography}

\end{document}